\documentclass[12pt]{article}
\usepackage{epsfig}
\usepackage{algorithm}
\usepackage{algorithmic}
\usepackage{latexsym,amsfonts,amsmath,amssymb}
\usepackage{graphicx}
\usepackage{bbm}
\usepackage{lscape}
\usepackage{url,epsf,color}
\usepackage{ulem}
\usepackage{caption}
\usepackage{amssymb}
\usepackage{epsfig}
\usepackage{amsmath}
\usepackage{amsfonts}
\usepackage{authblk}
\usepackage{url,epsf,color}
\usepackage{listings}
\usepackage{float}

\newcommand {\etal} {et~al.~}
\newcommand {\dn}[1] {\boldsymbol #1}

\begin{document}

\begin{center}
\begin{Large}
\noindent{\bf Bayesian spatio-temporal epidemic models with applications to sheep pox}\\
\end{Large}
\end{center}

C. Malesios$^1$, N. Demiris$^2$, K. Kalogeropoulos$^3$ and I. Ntzoufras$^2$\\

\noindent\begin{small}{\it${}^1$Department of Agricultural Development, Democritus University of Thrace, Greece, Email: malesios@agro.duth.gr}\\
{\it ${}^2$Department of Statistics, Athens University of Economics and Business, Athens, Greece}\\
{\it ${}^3$Department of Statistics, London School of Economics, London, UK}\\
\end{small}

\begin{abstract}
{
Epidemic data often possess certain characteristics, such as the presence of many zeros, the spatial nature of the disease spread mechanism or
 environmental noise. This paper addresses these issues via suitable Bayesian modelling. In doing so we utilise stochastic regression models appropriate for spatio-temporal count data with an excess number of zeros. The developed regression framework can incorporate serial correlation and time varying covariates through an Ornstein Uhlenbeck process formulation. In addition, we explore the effect of different priors, including default options and techniques based upon variations of mixtures of $g$-priors. The effect of different distance kernels for the epidemic model component is investigated. We proceed by developing branching process-based methods for testing scenarios for disease control, thus linking traditional spatio-temporal models with epidemic processes, useful in policy-focused decision making. The approach is illustrated with an application to a sheep pox dataset from the Evros region, Greece.  }
\end{abstract}

\section{\label{sec1}Introduction}

In the present paper we develop suitable methodology for epidemic spatio-temporal data. Specifically, we extend previous analyses (e.g. Choi et al., 2012), by addressing several important aspects of epidemic modeling. In addition, we provide an epidemic interpretation of our modelling approach, thus gaining insights regarding potential future outbreaks.

Our modelling framework accommodates a number of features: First, we make use of spatial information related to the location of the infected premises. Specifically, we incorporate the spatial coordinates of each infected farm, allowing for the probability of infection between farms to depend upon their distance. A number of spatial transmission kernels are fitted to the spatio-temporal data using Bayesian methods and an extended investigation of their relative importance is performed. Second, we model the inherent serial correlation of the data generating mechanism via a latent Ornstein Uhlenbeck (OU) process evolving around a mean which is regressed upon a number of covariates, including the spatial transmission kernels. The mean of the OU process is allowed to vary across time, adapting to the changes of its predictors. Finally, we investigate one of the basic model selection problems in Bayesian regression-type modeling, namely the one of covariate selection. Various methods have been proposed in order to deal with this task (George and McCulloch, 1993; Kuo and Mallick, 1998). Here we tackle this issue by implementing recently developed variable selection methodology. The potential multicolinarity problems are accounted for via $g$-prior variations such as the unit information prior. For inference purposes we resort to Markov chain Monte Carlo (MCMC) simulation which offers flexibility in the ability to fit complex models of the kind entertained in this paper.

Sheep pox is a highly contagious viral infection that can have devastating consequences (Garner et al., 2000). The proposed methodology is applied to a historical dataset obtained from the sheep pox epidemic in the Evros Prefecture of Northeastern Greece which began on December 1994 and ended in December 1998, infecting $249$ premises. The overall number of dead animals was estimated at approximately $35,500$. The data comprised of daily records of infected herds, temporal information such as the day of culling, detection time of the virus, the putative infection time and farm-level data (i.e. type and number of animals in each farm). To avoid uncertainties related to the exact day of infection we analyze the aggregated weekly counts of infection. A preliminary investigation of the data is presented in Malesios \etal  (2014). In particular, it appears that the sheep pox data strongly support the Poisson and zero inflated Poisson (ZIP)-based models as opposed to those based on the negative binomial distribution. Hence, in this paper we focus upon the Poisson and ZIP specifications.

The following section contains the model formulation while section \ref{secEpiCon} presents the connection of this class of models with stochastic epidemics, a useful tool for disease control. Section $4$ illustrates the results of the analysis and section $5$ concludes with some discussion.

\section{\label{secModel}Bayesian modelling}
\subsection{Spatio-temporal model}

The data consist of $260$ weekly observations spanning over a time interval of $5$ years. Time is measured in weeks and we let $t_i$, $\, i \in \{0,1,\dots,259\}$ correspond to the time after the $i$-th week. The response vector $\dn{y}=\{ y_i \, ; \, i=0, \dots, 259 \}$ is also ordered chronologically where each $y_{i}$ denote the number of sheep pox cases at $t_i$. The fitted models are special cases of the specification scheme described by:
\begin{equation}
\begin{cases}
y_i &\sim  g(y_i|\theta_i,p_i)  \\
g(y_i|\theta_i,p_i) &= p_i I_{\{y_i=0\}}+(1-p_i)f(y_i|\theta_i)  \\
\theta_i &= h(\lambda_i)=\exp(\lambda_i) \\
d\lambda_{t} &= \phi(\lambda_t-\mu_t)dt + dB_t \\
\end{cases}\label{model_formulation}
\end{equation}
where $B_t$ denotes standard Brownian motion, and $\mu_t$ is the piecewise constant deterministic process:
\begin{equation*}
\mu_t=
\begin{cases}
\mu^{(0)} & \text{if} \;\; 0\leq t < t_1 \\
\mu^{(1)} & \text{if} \;\; t_1\leq t <  t_2\\
\vdots & \\
\mu^{(259)} & \text{if} \;\; t_{259}\leq t < t_{260}
\end{cases}
\end{equation*}
where each $\mu^{(i)}$ corresponding to $t_i\leq t<t_{i+1}$, $i=0,1, \dots, 259$, is given by
\begin{eqnarray*}
\mu^{(i)} & = & \mathbf{X}_{(i)}\dn{\beta}+b_{i}+\Theta_{\tau}y_{i-1} +
K \big( \mathsf{\bf d}_i,\mathbf{\Theta}_{K} \big).
\end{eqnarray*}

\noindent Moreover, $I_{\{y_i=0\}}$ is an indicator variable denoting whether the response is positive or not,
$\theta_i$ is the instantaneous rate of the process at the $i$-th time point, $p_i$ denotes the proportion of  excess zeros for $t_i$ and $\dn{\beta}$ is a vector of dimension $\dim(\dn{\beta})$ with the coefficients of the fixed-effects covariates;
the corresponding covariate values for $t_i$ are denoted by the row vector $\mathbf{X}_{(i)}$
[also of dimension $\dim(\dn{\beta})$].
In our dataset $\dim(\dn{\beta})=10$, and therefore $\mbox{{\boldmath$\beta$}}$=$(\beta_{0},\beta_{1},\beta_{2},\beta_{3},...,
\beta_{9})^T$ is the vector of regression coefficients for the
intercept $(\beta_0)$, the covariates describing the number of villages infected in the previous week $(\beta_1)$, rainfall
$(\beta_2)$, average temperature $(\beta_3)$, maximum temperature
$(\beta_4)$, minimum temperature $(\beta_5)$, average humidity
$(\beta_6)$ and seasonal effects: spring $(\beta_7)$, summer
$(\beta_8)$, autumn $(\beta_9)$. \
In addition,  $\mathbf{\Theta_{K}}$ is the parameter vector of the transmission kernel function $K( \cdot )$ and $b_i$ reflects independent yearly random effects with $\dn{b} = (b_0, b_1, b_2, b_3, b_4)^T$. Note that Poisson regression is recovered when $p_i=0$ and $f$ denotes the Poisson probability mass function.

The instantaneous $\lambda_t$ is an Ornstein-Uhlenbeck process evolving around $\mu_t$, which in turn is determined by the potentially time varying covariates $\mathbf{X}_{(i)}$ and the spatial transmission kernels. Its transition density is available in closed form allowing us to write (for all $i$)
$$
\lambda_{t_{i+1}}|\lambda_{t_i} \;
\sim N \left( \mu^{(i)}+\left(\lambda_{t_i}-\mu^{(i)}\right)e^{ -\phi \delta_{i}}, \;
                      \dfrac{1-e^{-2\phi\delta_i}}{2\phi}  \right),\;\;\delta_i=t_{i+1}-t_i.
$$
The OU process reflecting $\lambda_t$ need not be stationary. In
fact, every change in the covariates provides a shock to the
system, to which the latent process $\lambda_t$ adapts through a
transient OU process with rate of convergence driven by $\phi$.
The Brownian motion $B_t$ is linked with the inherent
environmental noise of the system, while it can also absorb
potential model mis-specification. This formulation is therefore
substantially different than that of Branscum \etal (2008) and  Choi \etal  (2012), where the
covariance of the process is also modeled via that of a stationary
OU process in the spirit of Taylor \etal  (1994). It may actually
be seen as the model used in Struthers and McLeish (2011), adapted
to the context of this paper. We adopt this formulation as it is
seems more natural for our model given the various time varying
covariates. Nevertheless, it is still possible to fit a model
based on a zero-mean stationary OU component, see Table S2 for a comparison of the two approaches.

The $K \big( \mathsf{\bf d}_i, \mathbf{\Theta}_{K} \big)$ term is used to model  the spatial component of disease propagation; where $\mathsf{\bf d}_i = \{ \mathsf{d}_{k \ell}: k \in {\cal S}_{i}, \ell \in {\cal S}_{i-1} \}$,
is the set of all the Euclidean distances between farms $k \in {\cal S}_{i}$ and $\ell \in {\cal S}_{i-1}$,
and ${\cal S}_{i}$ is the set of farms with sheep pox incidence at time point $i$.
The Euclidean distance $d_{k \ell}$ is calculated by
$d_{k \ell} = \sqrt{ (u_k - u_l)^2 + (v_k - v_l)^2  }$
with $(u_k, v_k)$ denoting the geographical coordinates of farm $k$ measured in kms according to global positioning system (GPS). The geographical coordinates were then used for calculating the matrix containing all pairwise distances.

We may now complete our model formulation by using a similar structure for the zero-inflation probability $p_i$ as the one for $\lambda_i$ in (\ref{model_formulation}), writing
\begin{eqnarray}
\log \left( \frac{p_i}{1-p_i} \right) &=& \mathbf{X}_{(i)} \dn{\beta}^{z}+b^{z}_{t_i}+\Theta_{\tau}^{z}y_{i-1} +
K( \mathsf{\bf d}_i,\mathbf{\Theta}_K^z)~ ;
\end{eqnarray}
where the super-script $z$ denotes the parameters used for modeling the zero-inflation probability
with similar role as $\dn{\beta}$, $b_{t_i}$,  $\Theta_{\tau}$,  $\mathbf{\Theta}_K$, respectively.

\subsubsection{Spatial kernels}
Most of the attempts to capture the spatial structure of transmission of an animal disease have been performed for foot-and-mouth disease (FMD), due to its economically devastating consequences to livestock (Keeling, 2005). A large amount of epidemic data were collected during the 2001 UK FMD outbreak.
Thus, the use of simulation modeling for estimating the spread of highly contagious livestock diseases and for conducting risk assessment for various control measures has become common in recent years (e.g. Keeling et al., 2001, Tildesley et al., 2006, Chis-Ster and Ferguson 2007).

The $K \big( \mathsf{\bf d}_i, \mathbf{\Theta}_{K} \big)$ term measures the effect of the distance between infected farms. It seems reasonable to assume that the magnitude of disease transmission is negatively affected by distance. This relation is incorporated by including kernels, $K( \cdot )$, of the form:

\begin{equation*}
K \big( \mathsf{\bf d}_i, \mathbf{\Theta}_{K} \big) = \begin{cases}
\frac{1}{|\mathsf{\bf d}_i|} \sum \limits _{k \in {\cal S}_i} \sum \limits_{k \in {\cal S}_{i-1} }
{\cal K}( \mathsf{d}_{k \ell}, \mathbf{\Theta}_{K} ) \, &\text{if}\ \, y_i>0 \ \, \text{and}\ \,y_{i-1}>0\\[1.2em]
{\cal K}(1,\mathbf{\Theta}_{K} \big) \, &\text{if}\ \, y_i>0\ \, \text{and}\ \,y_{i-1}=0\\[0.5em]
{\cal K}(\mathsf{d}_{min}, \mathbf{\Theta}_{K} \big) \, &\text{if}\ \,  y_i=0 \
\end{cases},
\end{equation*}
where $|\mathsf{\bf d}_i|$ is the cardinality of $\mathsf{\bf d}_i$.

The pre-specified constant $\mathsf{d}_{min}$  denotes the minimum distance beyond which infections cannot occur
(see, e.g., Deardon et al., 2010). In our analysis we set $\mathsf{d}_{min} = 250km$, a distance sufficiently higher than the largest observed distance of $69$ kms occurred in the Evros prefecture. Finally, we use the distance of $1km$ in the case where we have occurrence of the disease in the current week,
and no occurrence at all in the previous week (Deardon et al., 2010),
implying that the distances  between previous and current week are approximately zero.

For specifying the transmission kernel we have resorted to a variety (see Table $\ref{postex2}$) of relevant functions.

\begin{center}
Table \ref{postex2} near here\\
\end{center}

\noindent In addition, we allow for a change in the parameter value at some time-point $t_{change}$ by using a latent indicator for $t_{change}$ and placing a Uniform(1, 260) prior on its range.

\subsubsection{Intensity Decomposition}\label{IntDecompos}

An interesting interpretation of such models can be found in Meyer et
al. (2012) where the model components are appropriately splitted,
disentangling the {\it epidemic and endemic} aspects of disease
dynamics. In particular, one can imagine that environmental
covariate information may well relate to the endemic part of the
disease while farm-to-farm contacts or number of villages infected
in the previous week are concerned with epidemic spread. Our model
can naturally be adapted to this framework via an additive
decomposition of $\mu_t$, the mean driving the instantaneous log rate of infection $\lambda_t$. The trajectory of $\mu_t$ can be split into its endemic and epidemic parts as
follows:
\begin{equation*}
\mu_t =
\mathbf{\Theta}_{endemic} +
\mathbf{\Theta}_{epidemic}
\end{equation*}
where $\mathbf{\Theta}_{endemic}$ ($\mathbf{\Theta}_{epidemic}$)
denotes the time-dependent endemic (epidemic) component.

This decomposition enhances our ability to inform control strategies since a large epidemic component (relative to the endemic component) would suggest imposing restrictions associated with the spatial allocation of farm structure in the region of interest, whereas the opposite results may indicate that most of infections are due to external factors and thus are less sensitive to such control measures (Brown et al., 2013).

\subsection{Prior specification for the variable selection component}

The variable selection problem arises naturally when one wishs to simplify the model structure by eliminating predictors with negligible effects (e.g. George and McCulloch, 1993). Bayesian variable selection typically involves the introduction of a vector of binary indicators $\dn{\gamma} \in \{ 0, 1 \}^{\dim(\dn{\beta})}$ which contains each possible combination of covariates to be included in the model. Then, MCMC methodology can be used to estimate the posterior distribution of $\dn{\gamma}$. The exploratory results of Malesios \etal  (2014) suggest that only a few of the variables under consideration, such as the number of villages infected in the previous week and certain meteorological/environmental variables should be included in the final model.

\subsubsection{Hyper g-prior setup.}

In the present analysis, we use the hyper-$g$ prior introduced by Liang \etal  (2008)
and implemented by  Bov\'{e} and Held (2011) in generalized linear models (GLMs).
All covariates are first centered and, following Ntzoufras \etal  (2003), we consider a slightly modified version of the hyper-g prior:
\begin{eqnarray}
f( \beta_0 ) & \sim & Normal( 0, 10^{4} ), ~~~~~
f( \dn{\beta}_{\setminus 0} | \beta_0, \sigma^2 )   \sim   Normal \Big( \dn{0}, g e^{\beta_0} (\dn{X}_{\setminus 0}^T \dn{X}_{\setminus 0} )^{-1} \Big)~,
\label{gprior}
\end{eqnarray}
where $\dn{\beta}_{\setminus 0}$ is the vector $\dn{\beta}$ excluding $\beta_0$, $\dn{X}_{\setminus 0}$ is the data matrix $\dn{X}$ without the column that corresponds to the intercept $\beta_0$ and $e^{\beta_0}$ denotes a rough estimate of $\lambda_i$ under the above prior setup.
This prior can be seen as the power prior of Ibrahim and Chen (2000) for the parameters
$\dn{\beta}_{\setminus 0}$ conditioned upon the constant $\beta_0$ and setting the power equal to $g$. Then the imaginary covariate values are equal to the observed ones while the imaginary responses are equal to $e^{\beta_0}$, the expected value under the null model (i.e. the simplest model). Hence, this prior accounts for $n/g$ additional data points.

The hyper-$g$ prior, further assigns a $Beta$ density to the shrinkage factor $g/(g+1)$ so that
$$
\frac{g}{1+g}\sim\ Beta \left( 1,\frac{\alpha}{2}-1 \right).
$$
We primarily focus on the hyper-$g$ prior using the value $\alpha=4$ which corresponds to the uniform prior on the shrinkage parameter but, as we demonstrate later in the paper, the results are fairly robust for different choices of  $\alpha \in (2,4]$.

\subsubsection{Sensitivity analysis and comparisons with other priors for variable selection.}

Bayesian variable selection is notorious for its sensitivity to the choice of prior, particularly the prior variances of $\dn{\beta}$ or its multiplicator $g$ in (\ref{gprior}). This is due to the well know Bartlett-Lindley paradox; see for details in Lindley (1957) and Bartlett (1957). To this end, we have performed a number of comparisons and sensitivity analyses to test for the robustness of our results. Specifically, we conducted the following analyses:
\begin{enumerate}
\item [a)] We tested for the sensitivity over different values of the hyper-parameter $a \in (2,4] $.

\item [b)] We compared our results with the hyper-g/n prior introduced by Liang \etal (2008) and suggested for
                        GLMs by Bov\'{e} and Held (2011) including similar sensitivity analysis as in (a).

\item [c)] We compared our results with the following default choices, previously reported in literature:
                        \begin{enumerate}
                        \item [i)] Zellner's g-prior with $g=n$, denoted by $ZG(n)$,

                        \item [ii)] Zellner's g-prior with $g=p^2$, denoted by $ZG(p^2)$,

                        \item [iii)] an empirical normal prior with an approximate unit information interpretation,
                                                    denoted by EIU.
                        \end{enumerate}
\end{enumerate}

\paragraph{Implementation Details.}
For both the hyper-g and the hyper-g/n hyper-priors, we perform
variable selection using a variety of values for hyper-parameter $\alpha$
and we graphically examine the robustness of the posterior inclusion probabilities for each covariate.
We expect that results will be robust as previously reported by Dellaportas \etal (2012).

Concerning the comparison with other priors, the $ZG(n)$ can be thought as the default choice in Bayesian variable selection since it has a unit information interpretation and its results correspond asymptotically to those obtained via BIC (Kass and Wasserman, 1995).
Following the modification of Liang \etal (2008), and in order to have prior structure equivalent to the hyper-g prior,
we use (\ref{gprior}) with $g=n$.
Similarly, we have also considered as an alternative the (modified) Zellner's g-prior (\ref{gprior})
with $g=p^2$ which corresponds to the risk inflation criterion of Foster and George (1994);
see also Fernandez \etal (2001) for a related comment.

Finally, following Ntzoufras (2009), we use as a rough yardstick an empirical independent prior with approximate unit interpretation (EIU).
It is comprised  by independent normal prior distributions for each $\beta_j$, i.e. $\beta_j \sim N( 0, n \sigma^2_{\beta_j} )$
with $\sigma_{\beta_j}$ set to the posterior standard deviation of each $\beta_j$ of the full model with flat priors.
This setup obviously uses information from the data to specify the prior variance but its multiplication with the sample size $n$
makes this effect minimal and approximately equivalent to one data-point in a similar manner to the $g=n$ choice.

\subsubsection{Prior distribution on model space.}

Concerning the prior specification of model indicators $\dn{\gamma}$, we primarily use the uniform prior on model space with $\gamma_j \sim \mbox{Bernoulli}(0.5)$. We also compare our results with the recently used beta-binomial prior on the model space where $\gamma_j \sim \mbox{Bernoulli}(p)$ with $p \sim \mbox{Beta}(1,1)$; see for example Chipman \etal (2001).
The latter is very useful in large scale problems (with large $p$) due to its shrinkage effect yielding parsimonious model structures.
Moreover, it allows for additional prior variability and robustness (Wilson \etal, 2010).

\subsubsection{Prior specification for the remaining parameters}

For the $\Theta_{\tau}$ parameter, associated with number of sheep pox cases in the previous week, a weakly informative normal prior was used with zero mean and large variance (equal to $10^4$). The same prior was assigned to the parameters of the vector $\mathbf{\Theta_{K}}$. Random yearly effects are assumed to follow a $N(0, \sigma_{b}^{2})$ density with $\sigma_b \sim U(0,100)$.

\section{\label{secEpiCon}Epidemic control}

This section is concerned with the connection of our model, which can be seen as a typical epidemiological model, to stochastic epidemic models, often considered as invaluable tools for disease control. One of the primary objectives of modelling the spread of an infectious disease is the ability to evaluate disease severity through key measures such as its reproduction number $R_0$, often interpreted as the average number of
infections produced by a single infective during their infectious period.
This facilitates for disease control via appropriate prophylactic measures. Specifically, one can evaluate the corresponding extinction probability $q$ as well as the effect of control strategies in achieving the {\it sine qua non} target of reducing $R_0$ below unity, thus securing that major outbreaks cannot occur. Here, we link our models to stochastic epidemics via considering the corresponding branching process. Subsequently we exploit this connection by exploring alternative, covariate-based, scenarios for the probability of hypothetical sheep pox outbreaks going extinct in the region of Evros.

A branching process represents an accurate approximation to a stochastic epidemic model (i) at the early stages of an outbreak when the number of infected individuals is much smaller than the population size and (ii) at the onset of disease re-emergence in the context of endemic diseases. Assuming constant $\lambda$ is probably reasonable in these two scenarios.

A general family for the offspring distribution $Z$ of branching processes is given by the power series family where:
\begin{equation*}
P(Z=r) = \alpha_{r} \frac{\lambda^r}{A(\lambda)}, \ \ A(\lambda) =
\sum_{r=0}^{\infty} \alpha_{r} \lambda^r,
\end{equation*}
with $\lambda$ being the canonical parameter and $\alpha_r\ge0$. Then the probability, $q(\lambda)$, of an epidemic going extinct is the smallest root of the equation: $A(q\lambda)=qA(\lambda)$; see for example Guttorp (1991) and Farrington \etal (2003). For $a_r= (r!)^{-1}$ we obtain the Poisson distribution whence $q(\lambda)$ can be numerically calculated as the smallest root of $\exp(q \lambda) = q \exp(\lambda)$.

We proceed by exploring the effect of particular covariates upon $q(\lambda)$. In particular, we utilise three distinct values for each covariate (minimum, median and maximum) keeping the other covariates fixed at their median values and for each covariate combination we simulate from the posterior density of $q(\lambda)$ by sampling from the posterior of the $\beta$'s. Note that by using the posterior output we preserve the correlation structure of the posterior density, an important aspect when estimating non-linear functionals such as $q(\lambda)$. For the ZIP model where $E(Y_i) = (1-p) \lambda + p \cdot0$, we adjust the extinction probability via $Pr(\text{extinction}) = 1 \wedge (q(\lambda) + p)$.

We also use a recent result due to Britton and Neal (2013) to estimate the expected time, say $E(A_{Q})$, the outbreak has Q infected farms via:
\begin{equation*}
E(A_{Q}) = \frac{\lambda^{Q-1}}{Q(1\vee\lambda)^{Q}} , Q = 1,2,...
\end{equation*}

This gives a somewhat complementary measure of disease propagation. The following section illustrates the application of the model and the control methods to real data.

\section{Application to sheep pox data}

Table $\ref{postex1}$ presents the number of sheep pox cases during the 1994-98 period in the Evros Prefecture, Greece. We first illustrate the variable selection procedure through detailed comparisons and extensive sensitivity analyses (Section 4.1.1). Then we investigate the form of the spatial kernels (Section 4.1.2).
Having selected our model, we disentangle the endemic and epidemic components in Section 4.2 and estimate the extinction probabilities under several covariate scenarios in Section 4.3.

\begin{center}
Table \ref{postex1} near here\\
\end{center}

\subsection{Model building and choice of variables}

For all MCMC runs we used an output of ten thousand iterations produced from
chains with total length equal to 105,000 iterations and after
using a burn-in of 5,000 and a thinning lag of 10 iterations. The analyses were conducted using the WinBUGS software (Lunn et al., 2000). The codes are available in Appendix B of the supplementary materials.

In the following analysis, we have used  the Struthers and McLeish-like OU structure. Those models gave a
substantially better fit to the data (see Table S1 in Appendix A) and their running times were about five times lower than those of the Taylor-based alternatives.

\subsubsection{Covariate selection}

\paragraph{Sensitivity analysis.}

\begin{figure}[H]
\centerline{\epsfig{figure=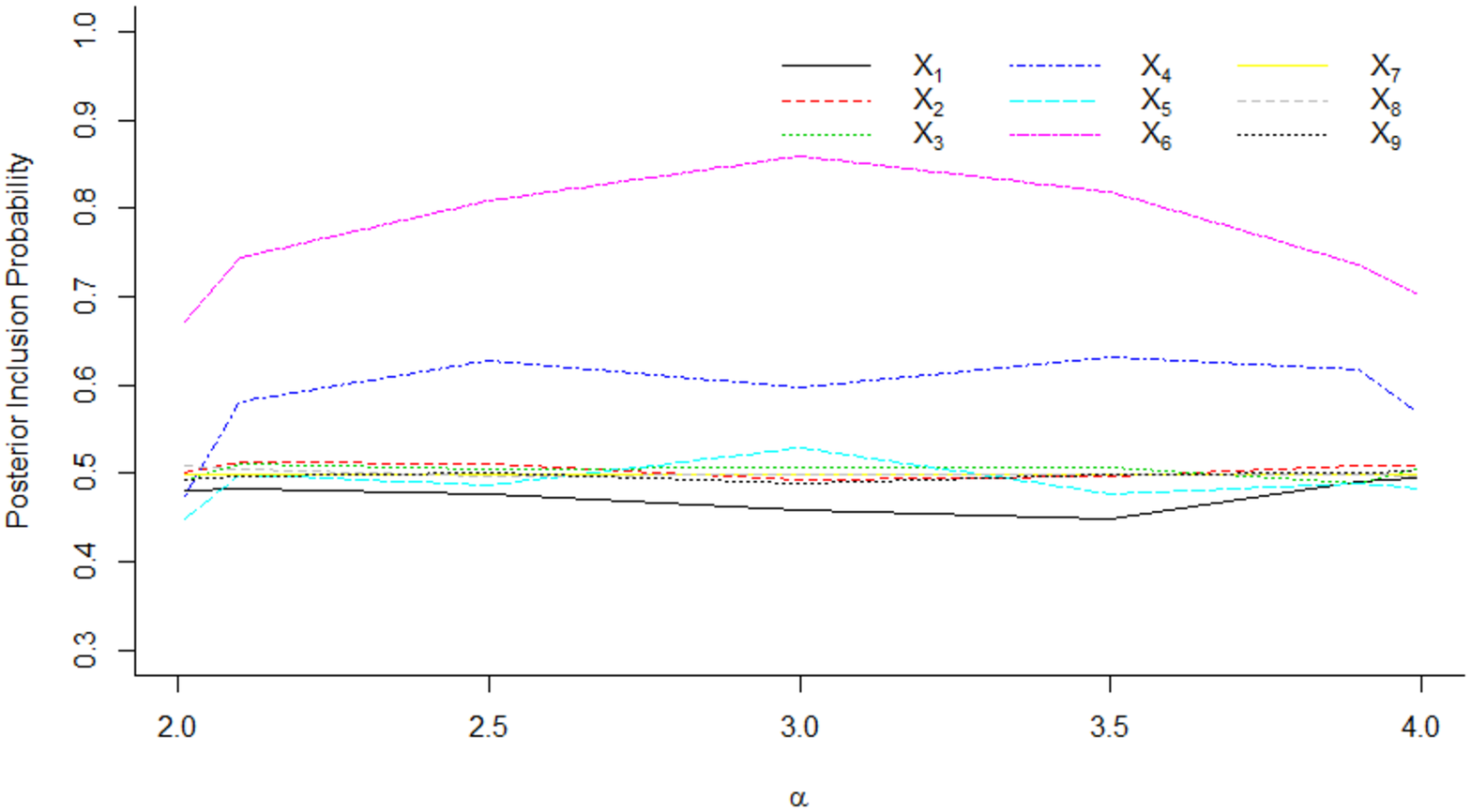,height=8cm,width=11cm}}
\caption{Sensitivity analysis of posterior inclusion probabilities for each covariate of infection rate $\lambda_{t}$ of the hyper-$g$ (uniform prior on model space). }\label{fig1:eps}
\end{figure}

\begin{figure}[H]
\centerline{\epsfig{figure=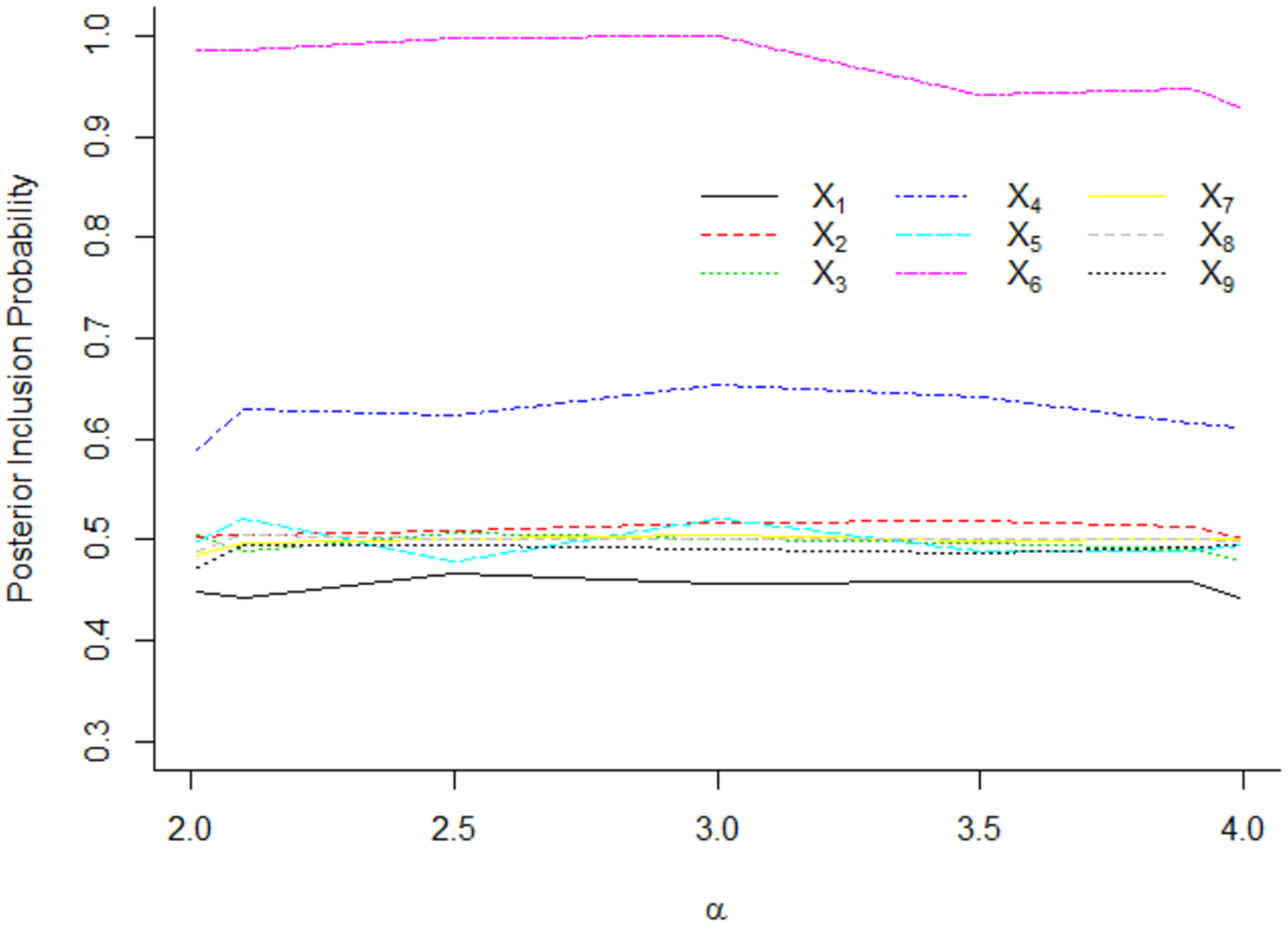,height=8cm,width=11cm}}
\caption{Sensitivity analysis of posterior inclusion probabilities for each covariate of infection rate $\lambda_{t}$ of the hyper-$g/n$ (uniform prior on model space). }\label{fig2:eps}
\end{figure}

We performed sensitivity analyses using the hyper-parameter values $\alpha \in \{ 2.01, 2.1, 2.5, 3.0, 3.5, 3.9, 3.99\}$
for the hyper-$g$ and the hyper-$g/n$ prior setups. The results are summarized in Figures $\ref{fig1:eps}$ and $\ref{fig2:eps}$ which present posterior inclusion probabilities under the hyper-$g$ and the hyper-$g/n$ priors, respectively, for each covariate regressed on the rate of infection.
The corresponding results for the covariates related to the probability of excess zeros are depicted in Figures S1 and S2 in Appendix A. These analyses have been conducted using the uniform prior on model space. The results obtained using the beta-binomial prior (not shown), although quantitatively different to those of the uniform prior, display similar ordering of the importance of the covariates.

\begin{center}
Figure \ref{fig1:eps} near here\\
\end{center}

\begin{center}
Figure \ref{fig2:eps} near here\\
\end{center}

Comparing the outcomes of the analysis one may deduce that the results are reasonably robust, especially for the covariates associated with excess zeros. Hence, for subsequent analyses under a hyper-$g$ prior setup, we focus on the choice of $\alpha=4$ which corresponds to the uniform prior on the shrinkage parameter.

\paragraph{Comparisons with other priors}

\begin{figure}[H]
\centerline{\epsfig{figure=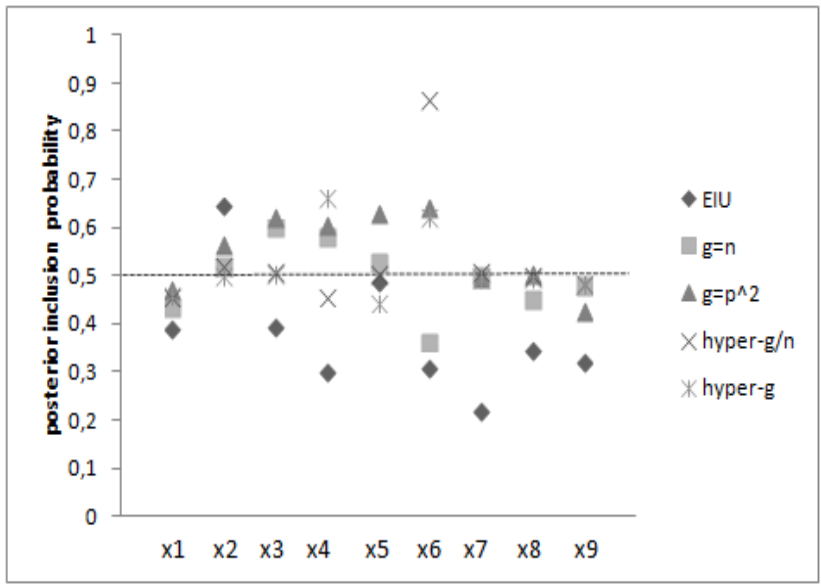,height=8cm,width=11cm}}
\caption{Inclusion probabilities for each covariate of infection rate $\lambda_{t}$ of the hyper-$g$ and comparison with other choices (uniform prior). }\label{fig3:eps}
\end{figure}

Figure $\ref{fig3:eps}$ presents a summary of the results on the comparison between the various choices of prior for covariate selection.

\begin{center}
Figure \ref{fig3:eps} near here\\
\end{center}

Five different priors are compared, notably EIU, $ZG(n)$, $ZG(p^2)$, hyper-$g$ and hyper-$g/n$ priors. In particular, Figure $\ref{fig3:eps}$ depicts posterior inclusion probabilities for each covariate of the infection rate for the hyper-$g$ prior and compares these values with the other choices. The corresponding results for the excess zeros are shown in Figure S3 in Appendix A. \\

The results suggest the inclusion of covariates $x_{4}$ and $x_{6}$, corresponding to the maximum temperature and the
average humidity and, potentially, the selection of covariates $x_{2}$ and $x_{3}$ (rainfall and average temperature, respectively) regarding the infection rate. On the other hand covariates $x_{5}$ and $x_{6}$ (i.e. minimum temperature and average humidity) are selected for the prediction of excess zeros. We decided to
keep for subsequent analyses the covariates with inclusion
probabilities over 0.6. These correspond to maximum temperature and humidity for the infection rate while minimum temperature and humidity are retained for the chance of excess zeros. In summary, it appears that a
combination of temperature and humidity seems largely responsible
for explaining disease occurrence. This is intuitively reasonable
and represents a common finding for animal diseases.

\paragraph{Final selection.}

For the remaining of this paper, we opt for the hyper-$g$ prior.
Hence, we focus on the results of the variable selection approach for the ZIP model under the hyper-$g$ prior specification (Table $\ref{postex3}$).

\begin{center}
Table \ref{postex3} near here\\
\end{center}

The results refer to the uniform prior on model space, however we have also run the model under the beta-binomial prior on model space to find a similar ordering for the covariates (see Table S1 in Appendix A), inflating upwards however the posterior inclusion probabilities in all covariates (almost all $\beta_{j}$ and $\beta^{z}_{j}$ - except for the covariates $x_{4}$, $x_{6}$ - ranged between 0.6 and 0.9). Table $\ref{postex4}$ includes the posterior summaries of the significant coefficients and the other parameters ($\phi$, random effects variance) for the model selected.

\begin{center}
Table \ref{postex4} near here\\
\end{center}

\subsubsection{Spatial kernels}

We proceed with fitting different kernel forms to the ZIP models in order to assess their relative importance in disease spread.  The posterior estimates for each $K$ parameter (i.e. $\alpha$, $\delta$ and/or $r$) are presented, especially $r$ which allows for occasional cases occurring far from the currently infected farms (Diggle, 2006).

\begin{figure}[H]
\centerline{\epsfig{figure=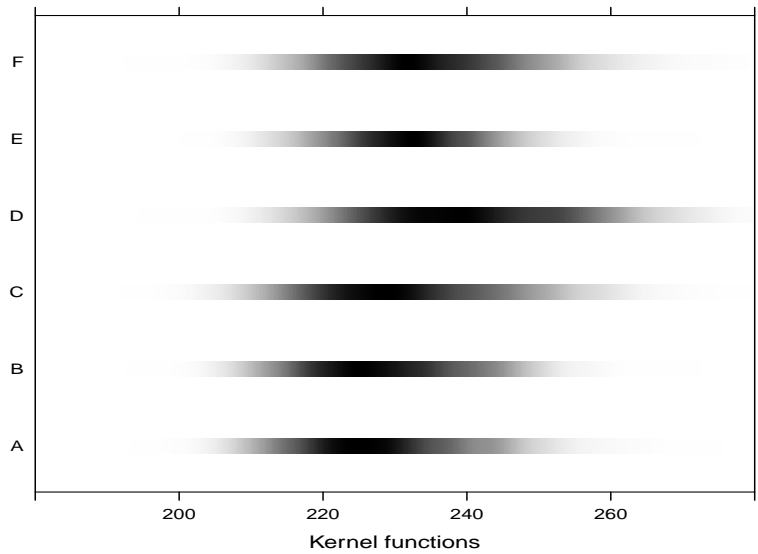,height=8cm,width=11cm}}
\caption{Posterior density strip plots of deviance $D$ for the six spatio-temporal models. }\label{fig4:eps}
\end{figure}

The inclusion of a change point in the kernel specification was deemed preferable to the alternative of no change point. Figure $\ref{fig4:eps}$ presents posterior density strip plots for the mean deviance $\overline{D}$ of the time-varying kernel models; see Table S2 (in Appendix A of the supplementary materials) for a comparison between the Struthers and McLeish (2011) and the Taylor \etal (1994) OU formulations. We additionally include the posterior mean deviance (denoted by $\overline{D^{'}}$) for the model without a spatial component.

\begin{center}
Figure \ref{fig4:eps} near here\\
\end{center}

For all the fitted kernels, the posterior distribution of the corresponding coefficients are well placed away from zero and the model fit is improved indicating the key importance of the spatial component in describing the progression of the epidemic.
There are mild differences in the ability of the models with different kernels to capture the observed dynamics of sheep pox occurrences. The fat-tailed kernel (A) (Chis-Ster and Ferguson 2007) yields the best fit, followed by the distance kernels (B) and (C). On the other hand, the distance kernels (D), (E) and (F) (e.g. Szmaragd et al.,  2009) gave a slightly worse fit. The better performance of the fat-tailed kernel may be an indication of the importance of long distances on disease spread when compared to the exponential-based functions (i.e. kernels B and C) which place less mass in the tails of the kernel functions. Thus, it appears that the spread of sheep pox epidemic in Evros was affected by long-range interactions. The coefficient $r$ (kernel C), measuring the relative importance of long-range transmission of sheep pox was also significant, confirming the above postulation. The change point is estimated to be around the $50^{th}$ week of the epidemic. The parameter $c$ is associated with the rate of decrease with distance. Higher values of $c$ indicate that disease incidence decreases faster with distance (local spread), whereas lower values indicate slower decrease. The current estimates of $c$ before and after the change point ($c_{pre}=8.26$ and $c_{post}=5.82$) suggests that during the first year ($1995$) the distance between farms of previous and current week was not especially significant in the limited – up to this time - spread of the disease. This period relates to sparse disease occurrence. In contrast, the disease subsequently displayed periodic outbursts with slow decrease of the incidents with distance. Personal communication with the local authorities confirms that during 1996 the local veterinary services have  resorted to a policy change for confronting the disease. Specifically, they increased the imposed restrictions to farms located within a $3$km ring around a detected infection, by increasing the hitherto time limit of $21$ days to $45$ days, due to the incubation period of sheep pox in the region believed to be well beyond the $21$ days period. In summary, the association between spatial information and progression of the sheep pox disease appears to be best expressed by the following form:
\begin{equation*}
f(d_{k \ell})=\begin{cases}
(1+ \frac{d_{k \ell}}{7.98})^{8.26}, \, &if\ \, week\le50\\
(1+ \frac{d_{k \ell}}{11.13})^{5.82}, \, &if\ \, week>50
\end{cases}
\end{equation*}

\begin{figure}[H]
\centerline{\epsfig{figure=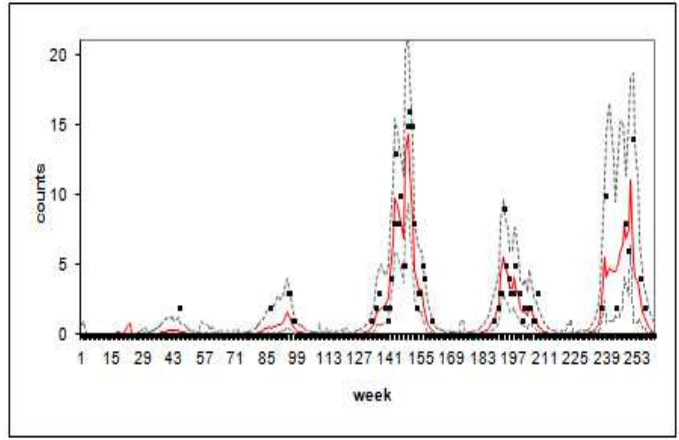,height=6cm,width=9cm}}
\caption{Predicted vs observed numbers of disease occurrence for the ZIP model. }\label{fig5:eps}
\end{figure}

Figure $\ref{fig5:eps}$ depicts the fit of the best model, indicating reasonably good agreement with the observed infectious disease counts.

\begin{center}
Figure \ref{fig5:eps} near here\\
\end{center}

\subsection{Endemic/epidemic decomposition}
The results regarding the endemic/epidemic decomposition presented
in Section $\ref{IntDecompos}$ are given in Table
$\ref{postex5}$ where the instantaneous mean ($\mu_t$) of the log
rate of infection ($\lambda_t$) is decomposed to its endemic
($\mathbf{\Theta}_{endemic} $) and epidemic
($\mathbf{\Theta}_{epidemic} $) components.

\begin{center}
Table \ref{postex5} near here\\
\end{center}

\begin{figure}[H]
\centerline{\epsfig{figure=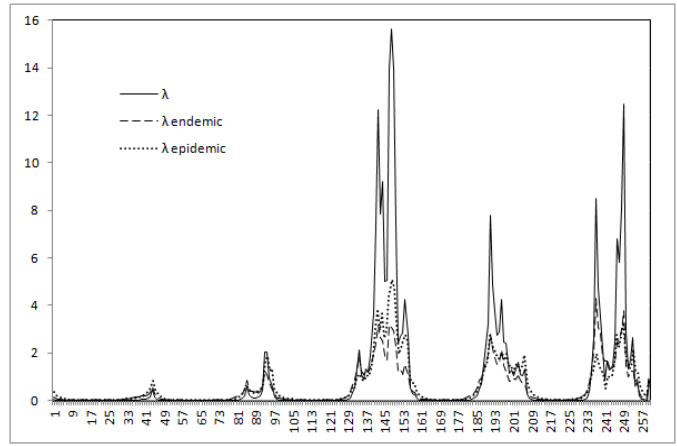,height=6cm,width=9cm}}
\caption{Epidemic and endemic decomposition of $\mu_{t}$ during the 1994-98 sheep pox epidemic in Evros Prefecture, Greece. }\label{fig6:eps}
\end{figure}

\begin{center}
Figure \ref{fig6:eps} near here\\
\end{center}

Specifically, we report the mean $\mu_t$ along with the
corresponding $95\%$ credible intervals under various scenarios
for a hypothetical outbreak. The results indicate that disease
spread is likely to increase for higher levels of the endemic
components. Figure $\ref{fig6:eps}$ demonstrates the
endemic/epidemic decomposition over the 5-year period (1994-1998)
of the sheep pox epidemic. This graph may assist in illustrating
the relative importance of the epidemic spatial component over the
endemic part of the model and vice versa during the progress of
the disease spread.

\subsection{Extinction Probabilities}

Here we present results based on the approach introduced in Section \ref{secEpiCon}, investigating the effect of each important covariate on a hypothetical future epidemic outbreak in the Evros region. Specifically, we combine parameter estimates from the historical sheep pox epidemic data and current farm locations in the region to calculate the probability of an epidemic going extinct. The findings are summarized in Table $\ref{postex6}$ and present extinction probabilities are obtained for the minimum, maximum and median value of each covariate, keeping the other covariates fixed at their median values (the covariate values used for the current analysis refer to year 2012 and are: min=-5.5, max=40.7, median=23.1 for maximum temperature; min=14, max=100, median=65 for humidity; min=0.02, max=45, median=16.26 for distance).

\begin{center}
Table \ref{postex6} near here\\
\end{center}

\begin{figure}[H]
\centerline{\epsfig{figure=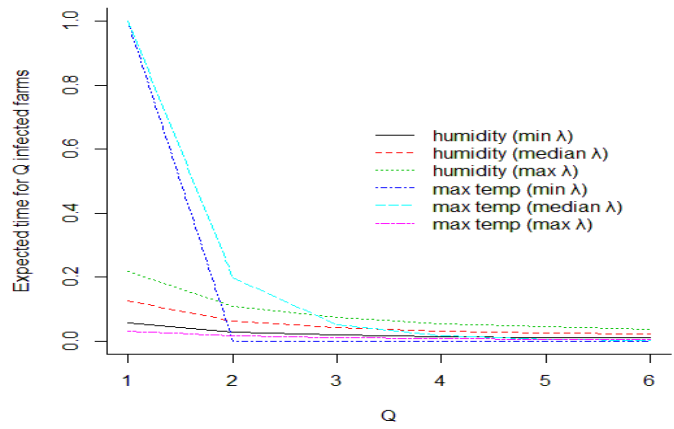,height=6cm,width=9cm}}
\caption{Expected time for exactly Q infected farms based on the branching process approximation.}\label{fig7:eps}
\end{figure}

It appears that a large epidemic may occur when (i) the distances between infected farms are small (average extinction probability, $q$=0.018), (ii) the levels of humidity are low ($q$=0.001) and (iii) the maximum temperatures are high ($q$=0.076). These results are based on the branching process approximation to the early stages of an outbreak and provide an indicator towards potential disease re-emergence. Therefore, monitoring these covariates may be useful for surveillance purposes.\\
The results for the expected `Q-occupation times', $Q \in \{1,2,3,4,5,6\}$, are summarized in Figure $\ref{fig7:eps}$, and are typical of a supercritical branching process (i.e. $\lambda_{t}>1$). Indeed, in the few occasions where $\lambda_{t}<1$ we expect only a few farms to get infected thus $E(A_{Q})$ is relatively large. In contrast, as $\lambda_{t}$ increases so does the chance of a large outbreak (see Table S3), leading to the apparent negative association between $\lambda_{t}$ and $E(A_{Q})$.

\begin{center}
Figure \ref{fig7:eps} near here\\
\end{center}

\section{Discussion}

In the present paper we proposed a general modelling framework which encompasses several common features of epidemic data. This was achieved by extending current spatio-temporal models via different variants of the O-U process. We conducted Bayesian variable selection by investigating recently developed priors which have not hitherto been used in models of high complexity. The extensive exploration of the spatial transmission component of the model suggests that a change-point appears necessary, offering a useful interpretation of the adopted policy. Perhaps more importantly, we developed methods to associate these spatio-temporal epidemiological models with stochastic epidemic processes through an approximate branching process representation. Thus, one can calculate suitable extinction probabilities under different scenarios providing a link to policy decisions targetted at disease control. In this paper we used standard Bayesian model determination techniques. An interesting alternative, that we intend to investigate in the future, can be based upon the prequential principle (Dawid, 1994).

\begin{center}
{\it Acknowledgments}
\end{center}
We are grateful to Zafeiris Abas and Omiros Papaspiliopoulos for helpful discussions.

\newpage

\noindent
{\bf References}

\medskip

\parskip 0.1cm

\noindent
{\bf Bartlett, M.S. (1957).} Comment on D.V. Lindley's Statistical Paradox. {\it Biometrika},  44, 533--534.

\noindent
{\bf Bov\'{e}, D.S. and Held, L. (2011).} Hyper-g priors for generalized linear models. {\it Bayesian Analysis}, 6(3), 387--410.

\noindent
{\bf Branscum, A.J., Perez, A.M., Johnson, W.O. and Thurmond, M.C. (2008).} Bayesian spatiotemporal analysis of foot-and-mouth disease data from the Republic of Turkey. {\it Epidemiology and Infection}, 136, 833--842.

\noindent
{\bf Britton, T. and Neal, P. (2013).} On the expected time a branching process has K individuals alive. {\it arXiv:1304.8014 [math.PR]}

\noindent
{\bf Brown, P.E., Chimard, F., Remorov, R., Rosenthal, J.S. and Wang, X. (2013).} Statistical inference and computational efficiency for spatial infectious-disease models with plantation data. {\it Journal of the Royal Statistical Society: Series C (Applied Statistics)} doi:10.1111/rssc.12036.

\noindent
{\bf Chipman, H., Ed George, McCulloch, R.E. (2001).} The Practical Implementation of Bayesian Model Selection (with discussion),
{\it IMS Lecture Notes - Monograph Series}, 38, 65 - 116

\noindent
{\bf Chis-Ster, I.C. and Ferguson, N.M. (2007).} Transmission parameters of the 2001 Foot and Mouth epidemic in Great Britain. {\it PLoS ONE}, 6, e502.

\noindent
{\bf Choi, Y.K., Johnson, W.O., Jones, G., Perez, A. and Thurmond, M.C. (2012).} Modelling and predicting temporal frequency of foot-and-mouth disease cases in countries with endemic foot-and-mouth disease. {\it Journal of the Royal Statistical Society A}, 175

\noindent
{\bf Deardon, R., Brooks, S.P., Grenfell, B.T., Keeling, M.J., Tildesley, M.J., Savill, N.J., Shaw, D.J. and Woolhouse, E.J. (2010).} Inference for individual-level models of infectious diseases in large populations. {\it Statistica Sinica}, 20, 239--261.

\noindent
{\bf Dellaportas P., Forster J.J. and Ntzoufras I. (2012).}
Joint Specification of Model Space and Parameter Space Prior Distributions. {\it Statistical Science}, 27, 232-246.

\noindent
{\bf Diggle, P.J. (2006).} Spatio-temporal point processes, partial likelihood, foot and mouth disease. {\it Statistical Methods in Medical Research}, 15, 325--336.

\noindent
{\bf Farrington, C.P., Kanaan, M.N. and Gay, N.J. (2003).} Branching process models for surveillance of infectious diseases controlled by mass vaccination. {\it Biostatistics}, 4(2), 279--295.

\noindent
{\bf Fernandez, C., Ley, E. and Steel, M.F. (2001).} Benchmark priors for Bayesian model averaging. {\it Journal of Econometrics}, 100, 381--427.

\noindent
{\bf Foster, D.P. and George, E.I. (1994).} The risk inflation criterion for multiple regression. {\it The Annals of Statistics}, 22, 1947--1975.

\noindent
{\bf Garner, M.G., Sawarkar, S.D., Brett, E.K., Edwards, J.R., Kulkami, V.B., Boyle, D.B. and Singh, S.N. (2000).} The extent and impact of sheep pox and goat pox in the state of Maharashtra, India. {\it Tropical Animal Health and Production}, 32(4), 205--223.

\noindent
{\bf George, E.I. and McCulloch, R.E. (1993).} Variable selection via Gibbs sampling. {\it Journal of the American Statistical Association}, 88(423), 881--889.

\noindent
{\bf Guttorp, P. (1991).} {\it Statistical inference for branching processes.} New York: Wiley.

\noindent
{\bf Ibrahim, J.G. and Chen, M.H. (2000).} Power Prior Distributions for Regression Models, {\it Statistical Science,}, 15 46--60.

\noindent
{\bf Kass, R.E. and Wasserman, L. (1995).} A reference Bayesian test for nested hypotheses and its relationship to the Schwarz criterion. {\it Journal of the American Statistical Association}, 90, 928--934.

\noindent
{\bf Keeling, M.J., Woolhouse, M.E., Shaw, D.J., Matthews L., Chase-Topping, M., Haydon, D.T., Cornell, S.J., Kappey, J., Wilesmith, J. and Grenfell, B.T. (2001).} Dynamics of the 2001 UK foot and mouth epidemic: stochastic dispersal in a heterogeneous landscape. {\it Science}, 294, 813--817.

\noindent
{\bf Keeling, M.J. (2005).} Models of foot-and-mouth disease. {\it Proceedings of the Royal Society, B, Biological Sciences}, 272(1569), 1195-1202.

\noindent
{\bf Kuo, L. and Mallick, B. (1998).} Variable selection for regression models. {\it Sankhya}, B 60, 65--81.

\noindent
{\bf Liang, F., Paulo, R., Molina, G., Clyde, M.A. and Berger, J.O. (2008).} Mixtures of g priors for Bayesian variable selection. {\it Journal of the American Statistical Association}, 103(481), 410--423.

\noindent
{\bf Lindley, D.V. (1957).} A Statistical Paradox. {\it Biometrika} {\bf 44}, 187--192.

\noindent
{\bf Lunn, D.J., Thomas, A., Best, N. and Spiegelhalter, D. (2000).} WinBUGS - A Bayesian modelling framework: Concepts, structure, and extensibility. {\it Statistics and Computing}, 10, 325--337.

\noindent
{\bf Malesios, C., Demiris, N., Abas, Z., Dadousis, K. and Koutroumanidis, T. (2014).} Modelling Sheep pox Disease from the 1994-1998 Epidemic in Evros Prefecture, Greece. {\it Under revision}

\noindent
{\bf Meyer, S., Elias, J. and H\"{o}hle, M. (2012).} A space-time conditional intensity model for invasive meningococcal disease occurence. {\it Biometrics}, 68, 607--616.

\noindent
{\bf Ntzoufras, I. (2009).} {\it Bayesian Modeling Using WinBUGS}. Wiley Series in Computational Statistics, Hoboken, USA.

\noindent
{\bf Ntzoufras, I., Dellaportas, P. and Forster, J.J. (2003).} Bayesian variable and link determination for generalized linear models. {\it Journal of Statistical Planning and Inference}, 111(1-2), 165--180.

\noindent
{\bf Struthers, C.A. and McLeish, D.L. (2011).} A particular diffusion model for incomplete longitudinal data: application to the multicenter AIDS cohort study. {\it Biostatistics} 12(3), 493--505.

\noindent
{\bf Szmaragd, C., Wilson, A.J., Carpenter, S., Wood, J.L.N., Mellor, P.S. and Gubbins, S. (2009).} A modeling framework to describe the transmission of bluetongue virus within and between farms in Great Britain. {\it PLoS ONE}, 4(11), e7741.

\noindent
{\bf Taylor, J.M.G., Cumberland, W.G. and Sy, J.P. (1994).} A stochastic model for analysis of longitudinal AIDS data. {\it Journal of the American Statistical Association}, 89, 727--736.

\noindent
{\bf Tildesley, M.J., Savill, N.J., Shaw, D.J., Deardon, R., Brooks, S.P., Woolhouse, M.E., Grenfell, B.T. and Keeling, M.J. (2006).} Optimal reactive vaccination strategies for a foot-and-mouth outbreak in the UK. {\it Nature}, 440, 83--86.

\noindent
{\bf Wilson, M.A., Iversen, E.S., Clyde, M.A., Schmidler, S.C. and Schildkraut, J.M. (2010).} Bayesian model search and multilevel inference for snp association studies. {\it The Annals of Applied Statistics}, 4(3), 1342--1364.

\newpage

\bigskip
\begin{table}[H]
\centering
\begin{tabular}{llcl}
\hline
Notation & ${\cal K}(\mathsf{d}_{k \ell}, \mathbf{\Theta}_{K})$                         & $\mathbf{\Theta}_{K}$ & Reference \\
\hline
A  & $\Big(1+\dfrac{\mathsf{d}_{k \ell}}{a} \Big)^{-c}$                                 & $(a,c)$  &Chis-Ster and Ferguson (2007)\\[1.0em]
B  & $\exp \big\{-\big(\tfrac{\mathsf{d}_{k \ell}}{a} \big)^{c} \big\}$                 & $(a,c)$  &Keeling \etal  (2001) \\[1.0em]
C  & $\exp \big\{-\big(\tfrac{\mathsf{d}_{k \ell}}{a} \big)^{c} \big\}+r$               & $(a,c,r)$&Diggle (2006) \\[1.0em]
D  & $a \exp \left(-a \, \mathsf{d}_{k \ell} \right)$                                   & $a$      &Szmaragd \etal  (2009) \\[1.0em]
E  & $\dfrac{\alpha}{\sqrt\pi} \exp \left(-a^{2} \mathsf{d}^{2}_{k \ell} \right)$       & $a$      &Szmaragd \etal  (2009) \\[1.2em]
F & $\dfrac{a}{4} \exp \left(-a^{\frac{1}{2}} \mathsf{d}^{\frac{1}{2}}_{k \ell} \right)$& $a$      &Szmaragd \etal  (2009) \\
\hline
\end{tabular}
\caption{Summary of transmission kernel functions included in spatio-temporal models.}
\label{postex2}
\end{table}

\begin{landscape}
\begin{table}
\begin{tabular}{llllllllllll}
\hline
Month/year & 11/94 & 10/95 & 11/95 & 7/96 & 8/96 & 9/96 & 10/96 & 11/96 & 12/96 & 1/97 & TOTAL \\
\hline
Number of cases & 2 & 6 & 2 & 6 & 3 & 28 & 34 & 49 & 15 & 1 &  \\
\hline
Month/year & 8/97 & 9/97 & 10/97 & 11/97 & 12/97 & 7/98 & 8/98 & 9/98 & 10/98 & 11/98 & 249 \\
\hline
Number of cases & 11 & 20 & 10 & 7 & 5 & 12 & 2 & 14 & 14 & 6 &  \\
\hline
 &  &  &  &  &  &  &  &  &  &  &  \\
\end{tabular}
\caption{Number of sheep pox cases during period 1994-98 in Evros Prefecture, Greece by month }
\label{postex1}
\end{table}
\end{landscape}

\clearpage
\newpage

\begin{table}[H]
\centering
\begin{tabular}{lclc}
\hline
& \multicolumn{3}{c}{{Hyper-$g$-prior ($\alpha=4$)}} \\ \hline
{\small Parameter} & {\small $\gamma$} & {\small Parameter} & {\small $\gamma^{z}$}\\

$\beta_1$ & 0.494  &$\beta^{z}_1$ & 0.499 \\
$\beta_2$ & \textbf{0.508}  &$\beta^{z}_2$ & 0.489  \\
$\beta_3$ & \textbf{0.504}  & $\beta^{z}_3$& 0.486   \\
$\beta_4$ & \textbf{0.601}  & $\beta^{z}_4$& 0.114  \\
$\beta_5$ & 0.483 &  $\beta^{z}_5$& \textbf{0.600}  \\
$\beta_6$ & \textbf{0.625}  & $\beta^{z}_6$& \textbf{0.951}   \\
$\beta_7$ & 0.499  & $\beta^{z}_7$& 0.497  \\
$\beta_8$ & 0.497 &  $\beta^{z}_8$& \textbf{0.508}  \\
$\beta_9$& 0.486  & $\beta^{z}_9$& 0.484  \\ \hline
$\bar{D}$ & 222.9 &   &
\\ \hline
\end{tabular}
\caption{Mean probabilities of inclusion ($\gamma$ and $\gamma^{z}$) for the hyper-$g$-prior ($\alpha=4$) variable selection approach applied to ZIP model with flat prior on $\beta_{0}$ and uniform prior on model space (posterior inclusion probabilities above $50\%$ are indicated in bold).}
\label{postex3}
\end{table}

\begin{table}[H]
\centering
\begin{tabular}{lc}
\hline
Parameter & Estimates \\ \hline
$\beta_4$ (max temperature)& 0.015  \\
& (0.002,0.032)\\
$\beta_6$ (humidity)& -0.019 \\
&(-0.043,-0.011)\\
$\beta^{z}_5$ (min temperature)& 0.049 \\
&(0.006,0.133)\\
$\beta^{z}_6$ (humidity)& 0.054 \\
&(0.022,0.112)\\
$\phi$ & 2.171 \\
&(0.763,6.647)\\
$\sigma^{2}_{b_{i}}$& 0.201  \\
&(0.042,2.809)\\ \hline
\end{tabular}
\caption{Posterior medians and corresponding $95\%$ credible intervals of the ZIP model}
\label{postex4}
\end{table}

\clearpage
\newpage

\begin{table}[H]
\centering
\begin{tabular}{llll}
\hline
 & \multicolumn{1}{c}{min} & \multicolumn{1}{c}{median} & \multicolumn{1}{c}{max} \\
\hline
$\mathbf{\Theta}_{endemic}$ & \multicolumn{1}{c}{1.961} & \multicolumn{1}{c}{4.045} & \multicolumn{1}{c}{16.506} \\
 & \multicolumn{1}{c}{(1.904-2.017)} & \multicolumn{1}{c}{(4.021-4.068)} & \multicolumn{1}{c}{(16.309-16.703)} \\
$\mathbf{\Theta}_{epidemic}$ & \multicolumn{1}{c}{1.269} & \multicolumn{1}{c}{1.156} & \multicolumn{1}{c}{1.011} \\
 & \multicolumn{1}{c}{(1.256-1.282)} & \multicolumn{1}{c}{(1.145-1.166)} & \multicolumn{1}{c}{(1.01-1.013)} \\
\hline
\end{tabular}
\caption{Endemic/epidemic decomposition of $\mu_{t}$.}
\label{postex5}
\end{table}

\begin{table}[H]
\centering
\begin{tabular}{llll}

\hline
 &     \multicolumn{1}{c}{humidity} & \multicolumn{1}{c}{maximum temperature} & \multicolumn{1}{c}{distance}  \\
\hline
 &     \multicolumn{1}{c}{0.001}  & \multicolumn{1}{c}{0.878} & \multicolumn{1}{c}{0.018} \\
min &    \multicolumn{1}{c}{(0.0007-0.001)} & \multicolumn{1}{c}{(0.861-0.895)} & \multicolumn{1}{c}{(0.01-0.256)} \\
\hline
 &     \multicolumn{1}{c}{0.485} & \multicolumn{1}{c}{0.076} & \multicolumn{1}{c}{0.195} \\
max &   \multicolumn{1}{c}{(0.456-0.513)} & \multicolumn{1}{c}{(0.062-0.091)} & \multicolumn{1}{c}{(0.174-0.217)} \\
\hline
all covariates &     \multicolumn{1}{c}{0.173} & \multicolumn{1}{c}{} & \multicolumn{1}{c}{} \\
at median values &    \multicolumn{1}{c}{(0.153-0.194)} & \multicolumn{1}{c}{} & \multicolumn{1}{c}{} \\
\hline
\end{tabular}
\caption{Estimated average extinction probabilities ($q$) along with corresponding $95\%$ credible intervals based on the branching process approximation.}
\label{postex6}
\end{table}

\clearpage
\newpage

\newpage

\setcounter{table}{0}
\setcounter{figure}{0}
\renewcommand{\thetable}{S\arabic{table}}   
\renewcommand{\thefigure}{S\arabic{figure}}

\newpage

\noindent
\textbf{\Large Appendix A}

\begin{table}[H]
\centering
\begin{tabular}{lclc}
\hline
& \multicolumn{3}{c}{{Hyper-$g$-prior ($\alpha=4$)}} \\ \hline
{\small Parameter} & {\small $\gamma$} & {\small Parameter} & {\small $\gamma^{z}$}\\

$\beta_1$ & \textbf{0.632}  &$\beta^{z}_1$ & \textbf{0.651} \\
$\beta_2$ & \textbf{0.651}  &$\beta^{z}_2$ & \textbf{0.649}  \\
$\beta_3$ & \textbf{0.668}  & $\beta^{z}_3$& \textbf{0.638}   \\
$\beta_4$ & \textbf{0.821}  & $\beta^{z}_4$& 0.289  \\
$\beta_5$ & \textbf{0.637} &  $\beta^{z}_5$& \textbf{0.692}  \\
$\beta_6$ & \textbf{0.839}  & $\beta^{z}_6$& \textbf{0.93}   \\
$\beta_7$ & \textbf{0.653}  & $\beta^{z}_7$& \textbf{0.646}  \\
$\beta_8$ & 0.448 &  $\beta^{z}_8$& \textbf{0.642}  \\
$\beta_9$& \textbf{0.635}  & $\beta^{z}_9$& \textbf{0.636}  \\ \hline
$\bar{D}$ & 223.5 &   &
\\ \hline
\end{tabular}
\caption{Mean probabilities of inclusion ($\gamma$ and $\gamma^{z}$) for the hyper-$g$-prior ($\alpha=4$) variable selection approach applied to ZIP model with flat prior on $\beta_{0}$ and beta binomial prior on model space (posterior inclusion probabilities above $50\%$ are indicated in bold).}
\label{postexS1}
\end{table}

\begin{table}[H]
\centering
\begin{tabular}{lcc}
\hline
&  SM OU  & Taylor et al. OU\\
& process & process\\
\hline
Kernel & $\overline{D}$ & $\overline{D}$\\
\hline
A Chis-Ster and Ferguson (2007)& 228 & 274.8  \\
B Keeling et al. (2001)& 228.5  & 277.4\\
C Diggle (2006)& 231.3 & 279\\
D Szmaragd et al. (2009)& 240.4 & 283.7\\
E Szmaragd et al. (2009)& 231.9 & 281\\
F Szmaragd et al. (2009)& 234.6  & 287.4\\
\hline
& $\overline{D^{'}}$  \\
& 294.3  \\
\hline
\end{tabular}
\caption{Goodness-of-fit statistics for the time-varying spatial models.}
\label{postexS2}
\end{table}

\newpage

\begin{landscape}
\begin{table}
\centering
\begin{tabular}{lllllll|llllll}
\cline{2-13}
 & \multicolumn{12}{c}{$E(A_{Q})$} \\ 
\cline{2-13}
 & \multicolumn{6}{c|}{humidity} & \multicolumn{6}{c}{max temperature} \\ 
\cline{2-13}
 & \multicolumn{6}{c|}{Q} & \multicolumn{6}{c}{Q} \\ 
\cline{2-13}
 & 1 & 2 & 3 & 4 & 5 & 6 & 1 & 2 & 3 & 4 & 5 & 6 \\ 
\cline{2-13}
min $\lambda_{t}$ & 0.057 & 0.028 & 0.019 & 0.014 & 0.011 & 0.009 & 1 & 0.015 & 0 & 0 & 0 & 0 \\ 
\cline{1-1}
median $\lambda_{t}$ & 0.127 & 0.063 & 0.042 & 0.032 & 0.025 & 0.021 & 1 & 0.196 & 0.051 & 0.015 & 0.004 & 0.001 \\ 
\cline{1-1}
max $\lambda_{t}$ & 0.219 & 0.109 & 0.072 & 0.054 & 0.043 & 0.036 & 0.031 & 0.015 & 0.01 & 0.007 & 0.006 & 0.005 \\ 
\hline
\end{tabular}
\caption{Expected time for exactly Q infected farms, $E(A_{Q})$, for the various levels of covariates, based on the branching process approximation.}
\label{postexS3}
\end{table}
\end{landscape}

\newpage

\begin{figure}[H]
\centerline{\epsfig{figure=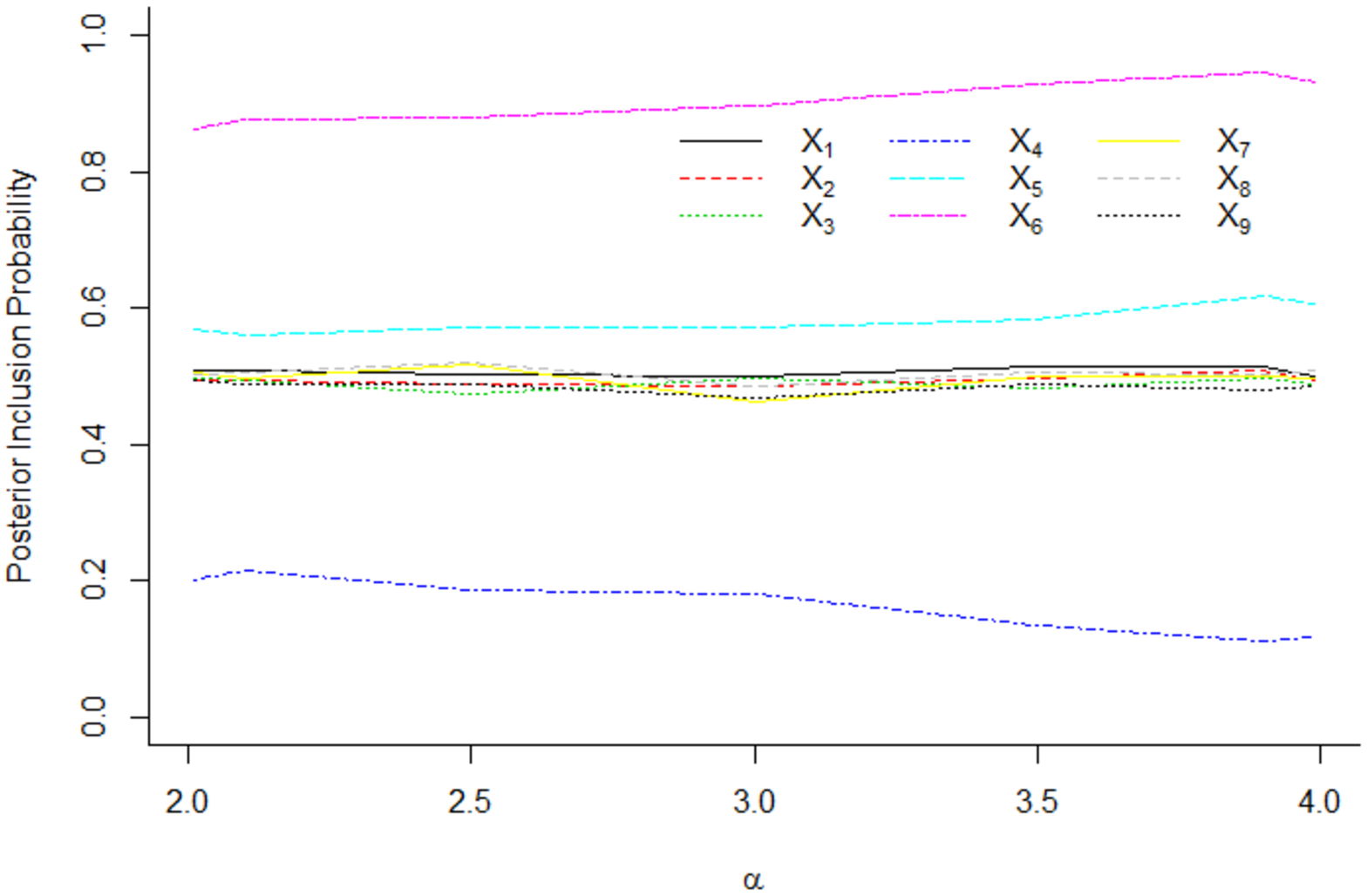,height=8cm,width=11cm}}
\caption{Sensitivity analysis of posterior inclusion probabilities for each covariate of excess zeros of the hyper-$g$ (uniform prior on model space). }\label{figS1:eps}
\end{figure}

\begin{figure}[H]
\centerline{\epsfig{figure=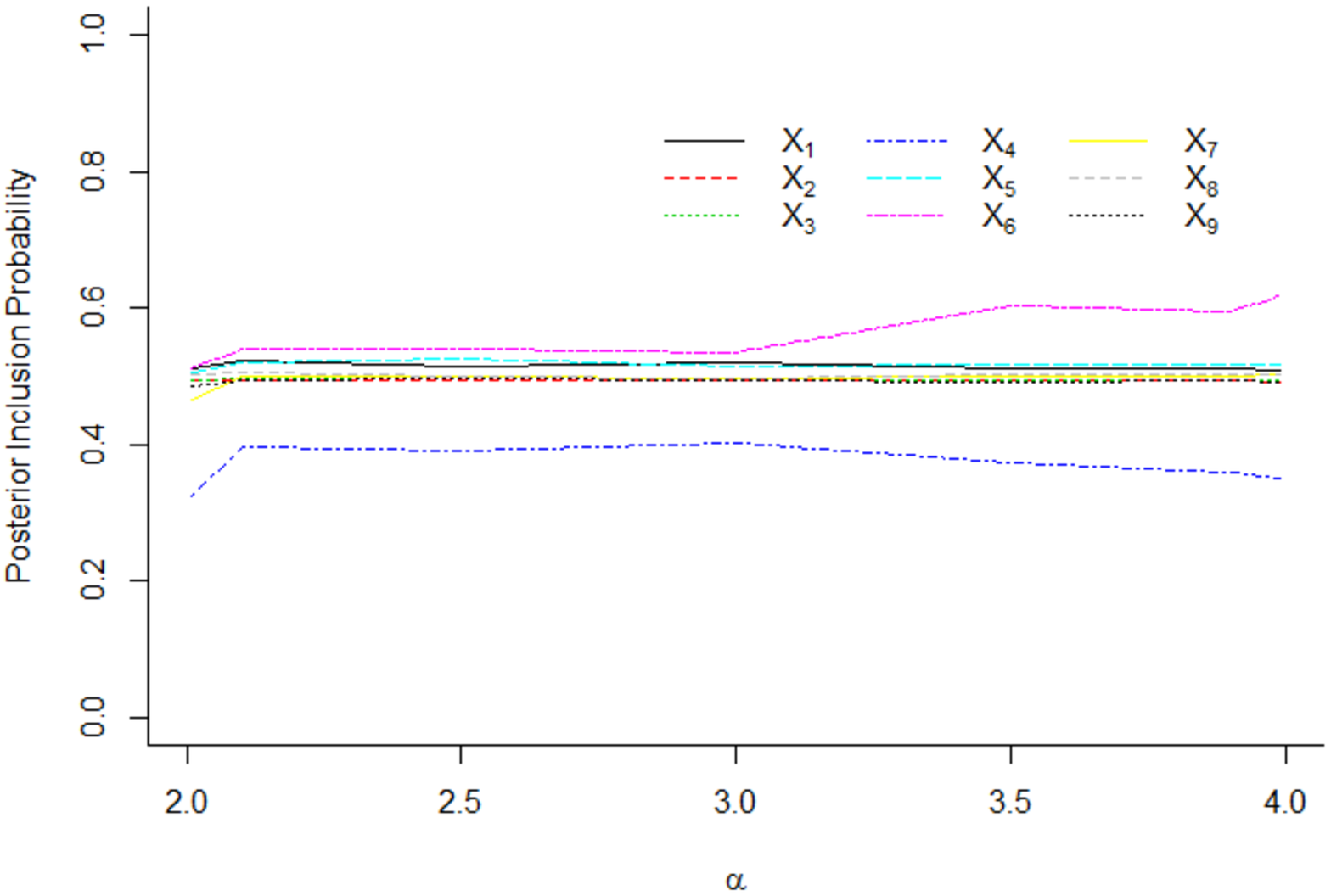,height=8cm,width=11cm}}
\caption{Sensitivity analysis of posterior inclusion probabilities for each covariate of excess zeros of the hyper-$g/n$ (uniform prior on model space). }\label{figS2:eps}
\end{figure}

\newpage

\begin{figure}[H]
\centerline{\epsfig{figure=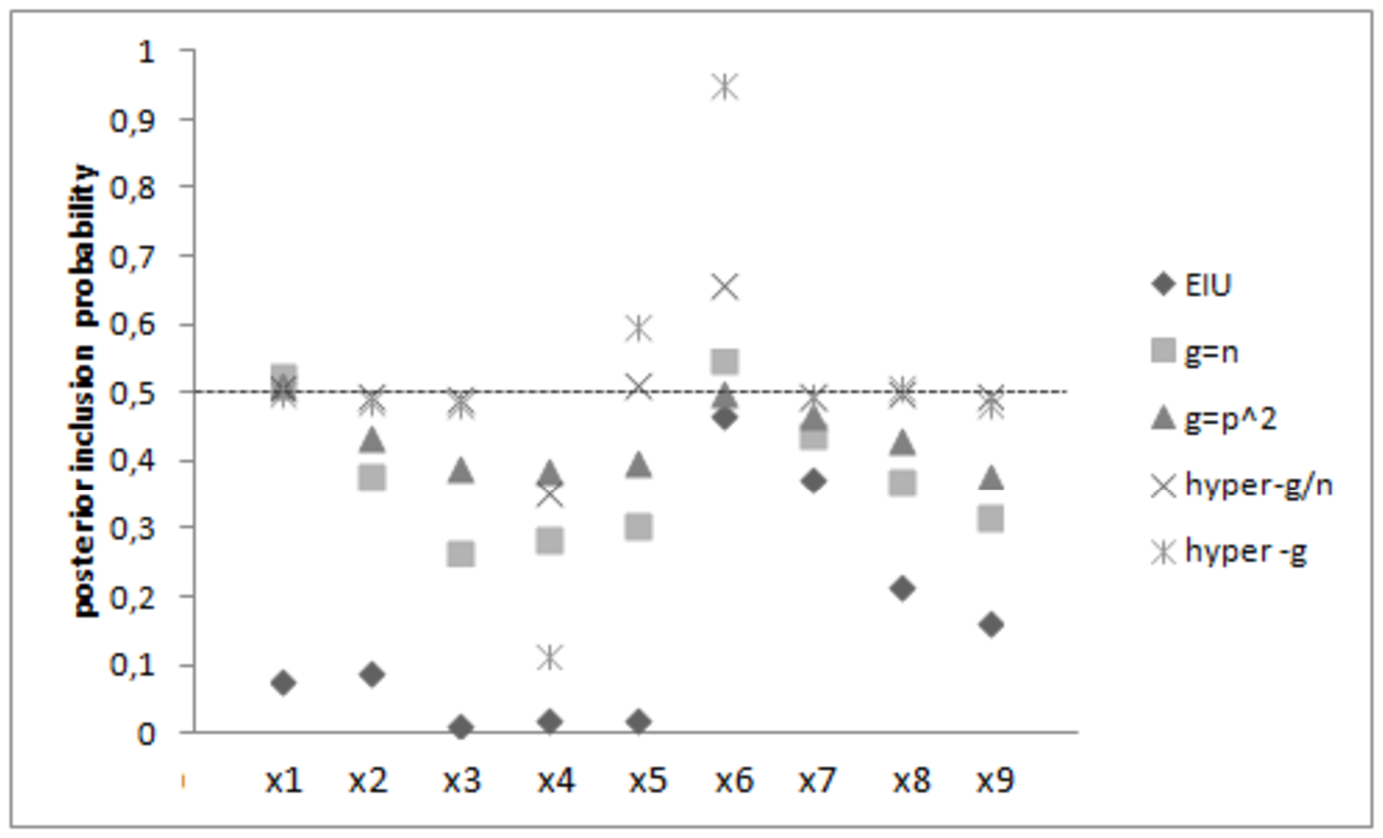,height=8cm,width=11cm}}
\caption{Inclusion probabilities for each covariate of excess zeros of the hyper-$g$ and comparison with other choices (uniform prior). }\label{figS3:eps}
\end{figure}

\newpage

\noindent
\textbf{\Large  Appendix B}

\noindent
\it{\#The WinBugs program for the best selected model (ZIP model with hyper-g prior, SM-OU specification, fat-tailed kernel (A))\#}

\begingroup
    \fontsize{10pt}{12pt}\selectfont
    \begin{verbatim}  

model
{
for (j in 1:9){  gb[j] <- b[j]*gamma1[j] } 
for (j in 1:9){  gc[j] <- c[j]*gamma2[j] } 

O[1] ~ dpois(lambda[1])
u[1] ~ dbern(p[1])
lambda[1] <- (1 - u[1]) * mu[1]
log(mu[1]) <- s + gb[1]*x1[1] +gb[2]*x2[1] +gb[3]*x3[1] + gb[4]*x4[1] + gb[5]*x5[1] 
+ gb[6]*x6[1] + gb[7]*sp[1] +gb[8]*su[1] + gb[9]*fa[1] 
+ b[10]*pow((1+(x10[1]/alpha1b[1])),-alpha2b[1]) + random[y[1]] 

logit(p[1]) <-s + gc[1]*x1[1] + gc[2]*x2[1] + gc[3]*x3[1] + gc[4]*x4[1] + gc[5]*x5[1] 
+ gc[6]*x6[1] + gc[7]*sp[1]+ gc[8]*su[1] + gc[9]*fa[1] 
+ c[10]*pow((1+(x10[1]/alpha1c[1])),-alpha2c[1])

#calculate each distance separately#
			for(k in 2:869) {
J[k] <- 1 + step(k - k.change)				
D1[k]<-pow((1+(x10[k]/alpha1b[J[k]])),-alpha2b[J[k]])
D2[k]<-pow((1+(x10[k]/alpha1c[J[k]])),-alpha2c[J[k]])
									}									
for(i in 2:n){

#sum of distances for each week#
V1[i]<-sum(D1[startinds[i]:endinds[i]])
V2[i]<-sum(D2[startinds[i]:endinds[i]])

O[i] ~ dpois(lambda[i])
u[i] ~ dbern(p[i])
lambda[i] <- (1-u[i]) * mu[i]
log(mu[i]) <-mu1[i]
mu1[i] ~ dnorm(M[i],U) 
C[i]<- s + gb[1]*x1[i]  + gb[2]*x2[i]  + gb[3]*x3[i] + gb[4]*x4[i] + gb[5]*x5[i] 
+ gb[6]*x6[i]  + gb[7]*sp[i] + gb[8]*su[i] + gb[9]*fa[i] + b[10]*V1[i] 
+ random[y[i]] + gam*O[i-1]

M[i]<-C[i] + (log(mu[i-1])- C[i])*exp(-phi) 

logit(p[i]) <-s + gc[1]*x1[i]  + gc[2]*x2[i] + gc[3]*x3[i] + gc[4]*x4[i] + gc[5]*x5[i] 
+ gc[6]*x6[i]  +gc[7]*sp[i] + gc[8]*su[i] + gc[9]*fa[i] + c[10]*V2[i]
}

k.change ~ dunif(3,259)

U<-(2*phi)/(1-exp(-2*phi))	
	
lamda<-exp(s)

#priors#
for (m in 1:5) {random[m] ~ dnorm(0, tau.btw)}
alpha1b[1] ~ dnorm( 0, 0.01)I(0, )
alpha2b[1] ~ dnorm( 0, 0.01)
alpha1b[2] ~ dnorm( 0, 0.01)I(0, )
alpha2b[2] ~ dnorm( 0, 0.01)

alpha1c[1] ~ dnorm( 0, 0.01)I(0, )
alpha2c[1] ~ dnorm( 0, 0.01)
alpha1c[2] ~ dnorm( 0, 0.01)I(0, )
alpha2c[2] ~ dnorm( 0, 0.01)
gam ~ dnorm(0, 0.01)
for (j in 1:9){ gamma1[j]~dbern(0.5) } 
for (j in 1:9){ gamma2[j]~dbern(0.5) } 


for (i in 1:10)
{
	for (j in 1:10)
		{
			inverse.V[i , j]<-inprod(x[ , i] , x[ , j])
		}
}


for (i in 1:9)
{
	for (j in 1:9)
		{
			prior.T[i , j]<-inverse.V[i , j]*lamda/(K/(1-K))
		}
}


s ~ dnorm( 0, 0.01)


b[1:9] ~ dmnorm( mu.beta[ ], prior.T[ , ])

for (j in 1:9)
	{mu.beta[j]<-0.0}


c[1:9] ~ dmnorm( mu.c[ ], prior.T[ , ])

for (j in 1:9)
	{mu.c[j]<-0.0}

K~dbeta(1,1)

for (k in 10:10) {b[k] ~ dnorm( 0, 0.01)}
for (k in 10:10) {c[k] ~ dnorm( 0, 0.01)}


phi<-exp(theta)
                        
theta ~ dnorm( 0, 0.01)


tau.btw ~ dgamma(0.1,0.1)
sbtw <- 1/tau.btw         

for (i in 1:n){ 
	x0[i]<-x[i,1]
	x1[i]<-x[i,2] 
	x2[i]<-x[i,3] 
	x3[i]<-x[i,4] 
	x4[i]<-x[i,5] 
	x5[i]<-x[i,6] 
	x6[i]<-x[i,7] 	
	sp[i]<-x[i,8] 
	su[i]<-x[i,9] 
	fa[i]<-x[i,10]}                  
}

#initial values#
list(theta=1,gam=0,tau.btw=1, b=c(0,0,0,0,0,0,0,0,0,0), c=c(0,0,0,0,0,0,0,0,0,0),
gamma1=c(0,0,0,0,0,0,0,0,0), gamma2=c(0,0,0,0,0,0,0,0,0),alpha1b=c(1,1), 
alpha2b=c(1,1),alpha1c=c(1,1), alpha2c=c(1,1),K=0.5,s=0)


data list( n=260,#Number of weekly disease occurrences#
O = c( 0, 0, 0, 0, 0, 0, 0, 0, 0, 0, 0, 0, 0, 0, 0, 0, 0, 0, 0, 0, 0, 0, 
0, 0, 0, 0, 0, 0, 0, 0, 0, 0, 0, 0, 0, 0, 0, 0, 0, 0, 0, 0, 0, 0, 2, 0, 0, 0, 
0, 0, 0, 0, 0, 0, 0, 0, 0, 0, 0, 0, 0, 0, 0, 0, 0, 0, 0, 0, 0, 0, 0, 0, 0, 0, 
0, 0, 0, 0, 0, 0, 0, 0, 0, 0, 0, 2, 0, 0, 0, 0, 0, 0, 0, 3, 3, 1, 1, 0, 0, 0, 
0, 0, 0, 0, 0, 0, 0, 0, 0, 0, 0, 0, 0, 0, 0, 0, 0, 0, 0, 0, 0, 0, 0, 0, 0, 0, 
0, 0, 0, 0, 0, 1, 0, 2, 3, 0, 0, 2, 1, 2, 4, 8, 13, 8, 10, 5, 5, 15, 16, 15, 8, 
2, 3, 3, 5, 4, 0, 0, 1, 0, 0, 0, 0, 0, 0, 0, 0, 0, 0, 0, 0, 0, 0, 0, 0, 0, 0, 
0, 0, 0, 0, 0, 0, 0, 0, 0, 1, 0, 2, 3, 3, 9, 5, 4, 3, 3, 5, 0, 3, 1, 2, 0, 2, 
2, 1, 1, 3, 0, 0, 0, 0, 0, 0, 0, 0, 0, 0, 0, 0, 0, 0, 0, 0, 0, 0, 0, 0, 0, 0, 
0, 0, 0, 0, 0, 0, 2, 0, 10, 0, 0, 0, 0, 2, 0, 0, 0, 8, 6, 0, 14, 0, 0, 0, 4, 0, 
2, 0, 0, 0, 0), 
#Year#
y = c( 1, 1, 
1, 1, 1, 1, 1, 1, 1, 1, 1, 1, 1, 1, 1, 1, 1, 1, 1, 1, 1, 1, 1, 1, 1, 1, 1, 1, 
1, 1, 1, 1, 1, 1, 1, 1, 1, 1, 1, 1, 1, 1, 1, 1, 1, 1, 1, 1, 1, 1, 1, 1, 2, 2, 
2, 2, 2, 2, 2, 2, 2, 2, 2, 2, 2, 2, 2, 2, 2, 2, 2, 2, 2, 2, 2, 2, 2, 2, 2, 2, 
2, 2, 2, 2, 2, 2, 2, 2, 2, 2, 2, 2, 2, 2, 2, 2, 2, 2, 2, 2, 2, 2, 2, 2, 3, 3, 
3, 3, 3, 3, 3, 3, 3, 3, 3, 3, 3, 3, 3, 3, 3, 3, 3, 3, 3, 3, 3, 3, 3, 3, 3, 3, 
3, 3, 3, 3, 3, 3, 3, 3, 3, 3, 3, 3, 3, 3, 3, 3, 3, 3, 3, 3, 3, 3, 3, 3, 4, 4, 
4, 4, 4, 4, 4, 4, 4, 4, 4, 4, 4, 4, 4, 4, 4, 4, 4, 4, 4, 4, 4, 4, 4, 4, 4, 4, 
4, 4, 4, 4, 4, 4, 4, 4, 4, 4, 4, 4, 4, 4, 4, 4, 4, 4, 4, 4, 4, 4, 4, 4, 5, 5, 
5, 5, 5, 5, 5, 5, 5, 5, 5, 5, 5, 5, 5, 5, 5, 5, 5, 5, 5, 5, 5, 5, 5, 5, 5, 5, 
5, 5, 5, 5, 5, 5, 5, 5, 5, 5, 5, 5, 5, 5, 5, 5, 5, 5, 5, 5, 5, 5, 5, 5) , 
#matrix of covariates#
x=structure(.Data=c( 1, 0, 0.76, 4.25, 5.2, 3, 87, 0, 0, 0, 1, 0, 3.39, 9.15, 12.229, 
5.3, 87, 0, 0, 0, 1, 0, 3.81, 9.657, 12.514, 6.314, 83.33, 0, 0, 0, 1, 0, 0.51, 
7.464, 9.743, 4.657, 83.67, 0, 0, 0, 1, 0, 0, 6.029, 10.057, 0.6, 85.67, 0, 0, 
0, 1, 0, 0, 6.136, 11.343, 0, 84.67, 0, 0, 0, 1, 0, 0.3428571, 6.55, 8.714, 
4.057, 89, 0, 0, 0, 1, 0, 2, 1.279, 5.486, -3.457, 85, 0, 0, 0, 1, 0, 0, 8.657, 
12.029, 4.657, 86, 0, 0, 0, 1, 0, 0.19, 10.443, 14.714, 4.886, 85.33429, 1, 0, 
0, 1, 0, 0, 7.293, 11.571, 2.4, 88.67, 1, 0, 0, 1, 0, 0, 9.921, 14.371, 2.971, 
77.33, 1, 0, 0, 1, 0, 2.07, 11.1, 15.6, 4.143, 85.67, 1, 0, 0, 1, 0, 4.6, 
9.321, 14.514, 3.629, 89.33, 1, 0, 0, 1, 0, 0, 14.05, 17.6, 9.2, 83.33, 1, 0, 
0, 1, 0, 2.057143, 15.557, 20.086, 7.886, 80.67, 1, 0, 0, 1, 0, 0.8, 16.121, 
20.571, 10.514, 82, 1, 0, 0, 1, 0, 0.26, 16.079, 20.029, 9.543, 83.85714, 1, 0, 
0, 1, 0, 0.1, 15.486, 19.457, 8.886, 85, 1, 0, 0, 1, 0, 0.33, 16.086, 19.514, 
8.629, 87.67, 1, 0, 0, 1, 0, 0, 21.221, 24.943, 11.2, 88, 1, 0, 0, 1, 0, 2.13, 
24.357, 28.343, 14.871, 77.67, 0, 1, 0, 1, 0, 0.11, 21.15, 25.914, 12.771, 
82.33, 0, 1, 0, 1, 0, 4.26, 20.664, 23.829, 13.133, 83.67, 0, 1, 0, 1, 0, 0, 
21.964, 26.229, 13.914, 66.33, 0, 1, 0, 1, 0, 0, 25.036, 28.943, 16.086, 
81.52429, 0, 1, 0, 1, 0, 0, 27.257, 32.543, 15.771, 73, 0, 1, 0, 1, 0, 7.04, 
27.6, 32.4, 18, 79.67, 0, 1, 0, 1, 0, 0, 23.864, 27.514, 17.571, 78, 0, 1, 0, 
1, 0, 0, 28.007, 32.143, 19.943, 84.67, 0, 1, 0, 1, 0, 0.01, 27.464, 31.771, 
19.343, 81.33, 0, 1, 0, 1, 0, 0, 26.871, 31.543, 18.057, 75.67, 0, 1, 0, 1, 0, 
1.11, 28.171, 32.771, 18.686, 80, 0, 1, 0, 1, 0, 1.1, 26.386, 31.086, 18.771, 
78.90571, 0, 1, 0, 1, 0, 0, 25.05, 30, 17.114, 53.67, 0, 1, 0, 1, 0, 0, 25.521, 
30.286, 18.257, 76.33, 0, 1, 0, 1, 0, 0, 25.329, 30.629, 17.4, 49.33, 0, 0, 1, 
1, 0, 0.07, 25.036, 31.686, 17.171, 74.33, 0, 0, 1, 1, 0, 0, 23.264, 29.829, 
14.457, 82, 0, 0, 1, 1, 0, 9.43, 23.571, 29.857, 16.114, 84.33, 0, 0, 1, 1, 0, 
2.26, 20.457, 25.657, 15, 82.33, 0, 0, 1, 1, 0, 0.6285714, 16.721, 20.029, 
13.171, 71.76, 0, 0, 1, 1, 0, 0.56, 15.179, 17.943, 12.571, 77.67, 0, 0, 1, 1, 
0, 0, 15.686, 20.257, 10.571, 75, 0, 0, 1, 1, 0, 3.942857, 14.443, 19.371, 
8.714, 76, 0, 0, 1, 1, 1, 4.73, 11.186, 14.371, 6.486, 67, 0, 0, 1, 1, 0, 0, 
8.779, 13.057, 4.971, 58.67, 0, 0, 1, 1, 0, 0, 7.286, 12.2, 2.743, 68.67, 0, 0, 
1, 1, 0, 0, 2.914, 8.571, -2.743, 70.67, 0, 0, 1, 1, 0, 1.36, 6.743, 11.429, 
-0.486, 70.52571, 0, 0, 0, 1, 0, 9.228571, 6.636, 12.257, 1.686, 67, 0, 0, 0, 
1, 0, 11.8, 4.607, 6.829, 1.486, 66, 0, 0, 0, 1, 0, 12.07143, 7.514, 10.686, 
5.657, 54.67, 0, 0, 0, 1, 0, 0.8714286, 3.793, 6, 2.4, 56.67, 0, 0, 0, 1, 0, 0, 
0.471, 2.4, -1.943, 62.67, 0, 0, 0, 1, 0, 0.2857143, 7.657, 12.543, 2.943, 
68.67, 0, 0, 0, 1, 0, 6.4, 6.357, 11.143, 2.543, 70.33, 0, 0, 0, 1, 0, 
0.6857143, 7.686, 11.571, 3.029, 63.71571, 0, 0, 0, 1, 0, 0, 8.171, 11.571, 
4.229, 87, 0, 0, 0, 1, 0, 0.5, 10.364, 15.057, 5.543, 83, 0, 0, 0, 1, 0, 
3.728571, 11.486, 15.8, 7.571, 74, 0, 0, 0, 1, 0, 1.86, 10.479, 14, 7.143, 
77.33, 1, 0, 0, 1, 0, 1.13, 6.55, 10.057, 2.343, 82.33, 1, 0, 0, 1, 0, 0.1, 
8.95, 13.114, 4.2, 90.67, 1, 0, 0, 1, 0, 14, 9.529, 13.314, 5.571, 79.33, 1, 0, 
0, 1, 0, 0, 12.95, 16.857, 6.571, 81.95143, 1, 0, 0, 1, 0, 4.73, 8.15, 12.8, 
3.571, 77, 1, 0, 0, 1, 0, 1.16, 13.607, 17.514, 6.686, 85.67, 1, 0, 0, 1, 0, 
1.46, 16.1, 20.257, 9.629, 85.67, 1, 0, 0, 1, 0, 0.8571429, 14.414, 18.886, 
8.6, 80.33, 1, 0, 0, 1, 0, 0, 18.736, 21.8, 11.2, 83.67, 1, 0, 0, 1, 0, 0, 
17.843, 21.857, 10.257, 52.67, 1, 0, 0, 1, 0, 0.4, 21.2, 25.371, 12.514, 51.67, 
1, 0, 0, 1, 0, 0, 23.743, 29, 13.229, 73.81143, 0, 1, 0, 1, 0, 0.04, 23.826, 
28.029, 16.143, 70.33, 0, 1, 0, 1, 0, 0.1, 26.736, 30.371, 17.429, 78.33, 0, 1, 
0, 1, 0, 1.74, 25.107, 29.943, 16.6, 85, 0, 1, 0, 1, 0, 0.1, 24.857, 28.314, 
17.657, 73.33, 0, 1, 0, 1, 0, 1.34, 26.093, 30.429, 17.343, 63.33, 0, 1, 0, 1, 
0, 2.07, 24.707, 29.457, 17.714, 62, 0, 1, 0, 1, 0, 0.09, 27.171, 31.314, 20.2, 
50.67, 0, 1, 0, 1, 0, 0, 27.829, 32.143, 19.086, 68.99857, 0, 1, 0, 1, 0, 0, 
26.193, 29.886, 18.657, 73.33, 0, 1, 0, 1, 0, 0.76, 24.843, 29.657, 16.857, 75, 
0, 1, 0, 1, 0, 0, 25.329, 29.4, 17.343, 88.33, 0, 1, 0, 1, 0, 1.31, 25.157, 
29.4, 17.2, 85.67, 0, 1, 0, 1, 1, 4.27, 22.943, 27.157, 15.457, 80.66, 0, 1, 0, 
1, 0, 0.6, 22.343, 26.114, 15.286, 69.33, 0, 1, 0, 1, 0, 0, 22.6, 27.229, 
16.086, 76.80857, 0, 0, 1, 1, 0, 0, 20.436, 24.686, 14.914, 71.67, 0, 0, 1, 1, 
0, 7.54, 17.6, 22.571, 11.2, 79.33, 0, 0, 1, 1, 0, 0, 15.15, 20.114, 8, 69, 0, 
0, 1, 1, 0, 0, 17.293, 22.2, 11.714, 65.33, 0, 0, 1, 1, 0, 0, 15.557, 20.6, 
9.057, 61.67, 0, 0, 1, 1, 2, 0.1, 11.943, 15, 9.371, 84.33, 0, 0, 1, 1, 1, 
18.42857, 12.907, 17.943, 7.143, 79.33, 0, 0, 1, 1, 1, 3.41, 4.786, 9.829, 
0.686, 72.95143, 0, 0, 1, 1, 1, 4.314286, 13.871, 17.086, 9.514, 78.67, 0, 0, 
1, 1, 0, 0.6, 2.714, 7.6, -0.971, 84, 0, 0, 1, 1, 0, 2.37, 8.907, 11.514, 
5.486, 57.33, 0, 0, 1, 1, 0, 0.57, 6.921, 8.857, 4.571, 55.33, 0, 0, 1, 1, 0, 
0, 4.6, 7.771, 0.4, 64.67, 0, 0, 0, 1, 0, 1.8, 6.364, 9.971, 1.714, 72.67, 0, 
0, 0, 1, 0, 4.585714, 11.757, 14.429, 9.8, 78.33, 0, 0, 0, 1, 0, 1.816667, 
5.767, 8.6, 2.9, 70.14286, 0, 0, 0, 1, 0, 0, 6.129, 8.2, 2.743, 82, 0, 0, 0, 1, 
0, 0, 1.764, 5.314, -0.971, 64, 0, 0, 0, 1, 0, 0.8857143, 3.007, 6.143, -0.457, 
69.33, 0, 0, 0, 1, 0, 0.27, 2.136, 5.2, -0.857, 87.33, 0, 0, 0, 1, 0, 2.03, 
3.193, 6.743, 0.457, 73.33, 0, 0, 0, 1, 0, 0.24, 4.471, 7.286, 0.914, 72.33, 0, 
0, 0, 1, 0, 14.04, 8.686, 11.943, 5.114, 80.33, 0, 0, 0, 1, 0, 3.085714, 3.136, 
6.543, -0.086, 75.52143, 0, 0, 0, 1, 0, 0.94, 2.129, 5.714, -1.914, 82.33, 1, 
0, 0, 1, 0, 0.9142857, 5.221, 7.829, 1.771, 79.67, 1, 0, 0, 1, 0, 1.06, 4.721, 
7.571, 1.057, 64.33, 1, 0, 0, 1, 0, 2.9, 8.357, 11.629, 4.229, 73.67, 1, 0, 0, 
1, 0, 4.428571, 12.2, 16.371, 7.114, 71.33, 1, 0, 0, 1, 0, 0, 10.707, 13.971, 
5.057, 70.33, 1, 0, 0, 1, 0, 3.685714, 10.021, 13.686, 4.8, 50.67, 1, 0, 0, 1, 
0, 0.03, 13.5, 17.971, 6.114, 70.33286, 1, 0, 0, 1, 0, 0, 18.057, 22.571, 
12.229, 49.33, 1, 0, 0, 1, 0, 0, 21.186, 26.743, 13.343, 68.33, 1, 0, 0, 1, 0, 
0.56, 19.35, 22.257, 13.343, 68.67, 1, 0, 0, 1, 0, 0, 22.514, 27.057, 15.543, 
69.33, 1, 0, 0, 1, 0, 1.51, 21.329, 25.029, 15.543, 69, 0, 1, 0, 1, 0, 0, 
23.457, 27.971, 14.4, 78, 0, 1, 0, 1, 0, 1.87, 25.371, 30.571, 16.286, 85.33, 
0, 1, 0, 1, 0, 0, 22.743, 26.514, 13.829, 69.71286, 0, 1, 0, 1, 0, 0, 25.55, 
29.8, 16.657, 83.67, 0, 1, 0, 1, 0, 0, 27.257, 31.8, 16.943, 81.67, 0, 1, 0, 1, 
0, 0, 26.779, 30.714, 18.629, 79.33, 0, 1, 0, 1, 1, 0, 27.014, 31.229, 19.029, 
70.67, 0, 1, 0, 1, 0, 0, 24.836, 28.571, 17.371, 58.33, 0, 1, 0, 1, 1, 0, 
28.05, 33.029, 19, 62.33, 0, 1, 0, 1, 1, 0.91, 26.293, 31.057, 18.171, 64.67, 
0, 1, 0, 1, 0, 0.04285714, 25.171, 29.543, 17.371, 71.52429, 0, 1, 0, 1, 0, 
0.7, 22.707, 26.486, 16.543, 80, 0, 1, 0, 1, 1, 0, 24.821, 28.8, 17.486, 80, 0, 
1, 0, 1, 1, 0.5714286, 21.85, 25.971, 15.8, 73.33, 0, 1, 0, 1, 2, 1.59, 19.921, 
23.743, 13.4, 52.67, 0, 0, 1, 1, 2, 0, 17.993, 22.657, 12, 72.67, 0, 0, 1, 1, 
5, 1.971429, 19.936, 23.829, 14.714, 66.67, 0, 0, 1, 1, 10, 0, 17.05, 20.886, 
12.857, 61, 0, 0, 1, 1, 6, 0.41, 16.414, 19.314, 12.857, 69.47714, 0, 0, 1, 1, 
8, 0, 15.214, 20.114, 9.171, 59.33, 0, 0, 1, 1, 4, 0, 12.586, 17.057, 5.971, 
80.67, 0, 0, 1, 1, 7, 2.19, 11.293, 16.486, 5.371, 61.33, 0, 0, 1, 1, 6, 0, 
11.829, 18.171, 4.629, 60, 0, 0, 1, 1, 8, 0, 9.207, 13.6, 4.829, 67.67, 0, 0, 
1, 1, 11, 2.54, 13.586, 16.486, 8.257, 75, 0, 0, 1, 1, 8, 14.25714, 12.471, 
15.771, 8.943, 75.33, 0, 0, 1, 1, 2, 6.31, 10.821, 12.914, 9.229, 68.47571, 0, 
0, 1, 1, 3, 0, 9, 11.343, 5.629, 69.33, 0, 0, 0, 1, 2, 0.7285714, 10.771, 14.4, 
7.286, 68.33, 0, 0, 0, 1, 4, 2.557143, 5.971, 9.629, 3.629, 66.33, 0, 0, 0, 1, 
2, 0, 8.575, 11.6, 3.85, 68.33, 0, 0, 0, 1, 0, 3.928571, 7.764, 10.029, 5.171, 
54.67, 0, 0, 0, 1, 0, 0.37, 4.264, 8.114, 1.2, 67.33, 0, 0, 0, 1, 1, 0, 5.543, 
10, 0.371, 66.67, 0, 0, 0, 1, 0, 0, 2.529, 7.971, -2.857, 65.85571, 0, 0, 0, 1, 
0, 0, 2.121, 7.343, -2.471, 67.67, 0, 0, 0, 1, 0, 0, 7.329, 12.771, 0.429, 
55.67, 0, 0, 0, 1, 0, 0.39, 5, 9.143, 0.629, 64, 0, 0, 0, 1, 0, 1.19, 7.85, 
11.971, 4.114, 75.33, 0, 0, 0, 1, 0, 0.3571429, 7.686, 12.429, 1.571, 64, 1, 0, 
0, 1, 0, 0.06, 7.064, 12.114, 1.371, 61.33, 1, 0, 0, 1, 0, 1.5, 7.321, 11.229, 
3.086, 48.67, 1, 0, 0, 1, 0, 0.04, 6.007, 9.629, 0.829, 62.38143, 1, 0, 0, 1, 
0, 7.642857, 9.036, 12.771, 5.914, 57.67, 1, 0, 0, 1, 0, 0, 8.107, 11.314, 2.6, 
59.67, 1, 0, 0, 1, 0, 4.76, 8.414, 11.829, 2.857, 68.33, 1, 0, 0, 1, 0, 0.84, 
12.979, 15.743, 8.829, 56.33, 1, 0, 0, 1, 0, 0.6571429, 13.607, 17.486, 7.771, 
65.67, 1, 0, 0, 1, 0, 0, 18.143, 22.143, 11.857, 63.67, 1, 0, 0, 1, 0, 0, 
22.421, 27.2, 14, 59, 1, 0, 0, 1, 0, 0.53, 21.557, 26.6, 12.886, 61.47714, 1, 
0, 0, 1, 0, 1.157143, 16.736, 19.857, 11.314, 74.67, 1, 0, 0, 1, 0, 0.1571429, 
19.429, 23.171, 11.857, 54, 0, 1, 0, 1, 0, 2.76, 22.364, 25.286, 15.371, 57.67, 
0, 1, 0, 1, 0, 0, 26.607, 30.057, 18.114, 61.67, 0, 1, 0, 1, 0, 0, 26.714, 
31.714, 18.257, 64.67, 0, 1, 0, 1, 0, 0, 28.043, 32.086, 18.057, 65, 0, 1, 0, 
1, 0, 0.1, 26.179, 29.371, 18.829, 56.67, 0, 1, 0, 1, 0, 1.3, 23.686, 27.343, 
15.343, 62.05, 0, 1, 0, 1, 0, 0, 26.479, 31.486, 18.914, 56.33, 0, 1, 0, 1, 0, 
1.01, 25.986, 31.2, 18.486, 67, 0, 1, 0, 1, 1, 5.714286, 24.521, 29.286, 
16.943, 58.67, 0, 1, 0, 1, 0, 0.4428571, 22.921, 28.2, 16.086, 81, 0, 1, 0, 1, 
2, 0, 23.157, 28.714, 16.2, 69, 0, 1, 0, 1, 1, 0.7857143, 22.643, 27.114, 10.1, 
62.67, 0, 1, 0, 1, 1, 0, 20.821, 26.371, 12, 67.67, 0, 1, 0, 1, 7, 0, 19.936, 
26.343, 11.243, 66.04857, 0, 1, 0, 1, 3, 0, 19.964, 25.657, 11.229, 66, 0, 0, 
1, 1, 3, 0, 16.593, 22.286, 8.4, 69.67, 0, 0, 1, 1, 3, 2.685714, 16.943, 
23.886, 9.829, 52, 0, 0, 1, 1, 2, 0, 16.914, 23.686, 6.357, 35, 0, 0, 1, 1, 3, 
2.04, 16.236, 20.557, 9.214, 49.67, 0, 0, 1, 1, 0, 5.771429, 11.164, 18.457, 
5.243, 53.33, 0, 0, 1, 1, 3, 0, 8.086, 12.4, 3.886, 58.33, 0, 0, 1, 1, 1, 0, 
9.786, 15.029, 3.386, 54.85714, 0, 0, 1, 1, 1, 3.285714, 14.621, 20.114, 
11.543, 43, 0, 0, 1, 1, 0, 11.74, 8.086, 11.914, 5.6, 45.67, 0, 0, 1, 1, 1, 0, 
10.836, 14.286, 7.6, 52.33, 0, 0, 1, 1, 1, 4.157143, 11.35, 14.8, 6.714, 51, 0, 
0, 1, 1, 1, 5.371429, 5.807, 9.6, 1.071, 40, 0, 0, 0, 1, 1, 2.41, 0.957, 6.6, 
-2.857, 38.67, 0, 0, 0, 1, 2, 2.86, 8.064, 12.286, 2.857, 37.33, 0, 0, 0, 1, 0, 
7.75, 6.45, 11, 0.733, 44, 0, 0, 0, 1, 0, 0, 6.857, 11.429, 2.143, 46.67, 0, 0, 
0, 1, 0, 0.01, 6.586, 10.457, 2.114, 53.33, 0, 0, 0, 1, 0, 5.53, 7.15, 10, 
4.371, 52.67, 0, 0, 0, 1, 0, 0, 1.564, 4.914, -2.314, 46.67, 0, 0, 0, 1, 0, 
11.81, 4.571, 7.229, 1.314, 54.33, 0, 0, 0, 1, 0, 0, 5.529, 10.029, 0.429, 
58.67, 0, 0, 0, 1, 0, 0.24, 7.793, 13.257, 2.343, 55.67, 0, 0, 0, 1, 0, 0, 
7.914, 14.229, 2.286, 52.57286, 0, 0, 0, 1, 0, 0, 8.986, 14.029, 1.386, 63.33, 
1, 0, 0, 1, 0, 9.1, 6.686, 12.914, 0.929, 65.33, 1, 0, 0, 1, 0, 1.057143, 
4.929, 9.557, 0.657, 79, 1, 0, 0, 1, 0, 0.33, 3.807, 6.743, 0.643, 73, 1, 0, 0, 
1, 0, 0.04, 9.243, 13.957, 3.2, 61.67, 1, 0, 0, 1, 0, 0, 16.807, 20.3, 7.929, 
51.33, 1, 0, 0, 1, 0, 0, 15.986, 20.029, 8.471, 41.67, 1, 0, 0, 1, 0, 0.24, 
13.414, 16.829, 7.771, 62.19, 1, 0, 0, 1, 0, 0.43, 16.271, 19.457, 10.886, 37, 
1, 0, 0, 1, 0, 0.2571429, 19.071, 23.343, 14.057, 42, 1, 0, 0, 1, 0, 3.2, 
16.714, 19.686, 13.143, 53, 1, 0, 0, 1, 0, 5.242857, 16.621, 19.143, 12.829, 
52, 1, 0, 0, 1, 0, 2.23, 20.443, 23.543, 14.314, 47.67, 1, 0, 0, 1, 0, 3.87, 
22.479, 26.057, 15.429, 46.67, 0, 1, 0, 1, 0, 0, 24.007, 26.857, 17.571, 44.33, 
0, 1, 0, 1, 0, 0.27, 23.557, 26.914, 15.571, 46.09571, 0, 1, 0, 1, 0, 
0.04285714, 25.229, 28.771, 16.229, 40.33, 0, 1, 0, 1, 0, 1.614286, 26.9, 31.8, 
19.257, 42.67, 0, 1, 0, 1, 0, 2.56, 22.371, 26.657, 15.343, 33.67, 0, 1, 0, 1, 
1, 0, 26.593, 30.571, 17.029, 35, 0, 1, 0, 1, 1, 0, 29.436, 34.029, 21.257, 
39.33, 0, 1, 0, 1, 1, 0, 30.4, 35.971, 20.943, 38, 0, 1, 0, 1, 0, 0, 31.771, 
36.457, 22.686, 38.33, 0, 1, 0, 1, 0, 0, 26.8, 31.6, 18.914, 38.19, 0, 1, 0, 1, 
0, 0, 25.8, 30.371, 18.686, 34, 0, 1, 0, 1, 0, 0, 25.929, 29.857, 18.886, 
40.67, 0, 1, 0, 1, 1, 0, 22.814, 26.771, 16.029, 45, 0, 1, 0, 1, 0, 0, 23.514, 
27.629, 17.371, 56.67, 0, 0, 1, 1, 0, 0.1428571, 20.3, 24.857, 14.286, 53, 0, 
0, 1, 1, 0, 3.67, 17.679, 21.371, 13.114, 39.67, 0, 0, 1, 1, 1, 4.79, 20.064, 
23.8, 14.743, 47, 0, 0, 1, 1, 1, 2.36, 19.171, 23.8, 13.343, 45.14429, 0, 0, 1, 
1, 0, 3.01, 16.943, 21.714, 12.714, 39.67, 0, 0, 1, 1, 1, 0, 15.421, 19.771, 
9.857, 53, 0, 0, 1, 1, 0, 8.728571, 13.864, 17.971, 9.6, 46, 0, 0, 1, 1, 0, 0, 
15.779, 19.457, 11.4, 49, 0, 0, 1, 1, 0, 6.014286, 9.386, 13.686, 5.343, 48, 0, 
0, 1, 1, 1, 4.242857, 7.229, 11.029, 3.686, 55.33, 0, 0, 1, 1, 0, 3.457143, 
9.286, 10.914, 7.457, 37, 0, 0, 1, 1, 1, 4.528571, 5.943, 7.829, 2.971, 
46.85714, 0, 0, 1, 1, 0, 2.3, 2.771, 9.2, 0.514, 43.33, 0, 0, 0, 1, 0, 0.29, 
4.5, 7.2, 1.314, 52, 0, 0, 0, 1, 0, 3.63, 3.557, 7.143, 0.343, 50.33, 0, 0, 0), 
.Dim = c(260,10)),
#distances#
x10 = c( 250, 250, 250, 250, 250, 250, 250, 250, 250, 250, 
250, 250, 250, 250, 250, 250, 250, 250, 250, 250, 250, 250, 250, 250, 250, 250, 
250, 250, 250, 250, 250, 250, 250, 250, 250, 250, 250, 250, 250, 250, 250, 250, 
250, 250, 1, 250, 250, 250, 250, 250, 250, 250, 250, 250, 250, 250, 250, 250, 
250, 250, 250, 250, 250, 250, 250, 250, 250, 250, 250, 250, 250, 250, 250, 250, 
250, 250, 250, 250, 250, 250, 250, 250, 250, 250, 250, 1, 250, 250, 250, 250, 
250, 250, 250, 1, 20.31, 18.68, 8.383, 9.022, 20.52, 250, 250, 250, 250, 250, 
250, 250, 250, 250, 250, 250, 250, 250, 250, 250, 250, 250, 250, 250, 250, 250, 
250, 250, 250, 250, 250, 250, 250, 250, 250, 250, 250, 250, 250, 1, 250, 1, 
20.6, 250, 250, 1, 12.74, 17.42, 18.62, 11.83, 4.117, 6.24, 11.83, 4.073, 
8.228, 9.884, 2.021, 10.39, 4.117, 10.51, 8.918, 8.761, 6.145, 11.89, 17.63, 
14.2, 27.3, 18.06, 6.24, 2.563, 4.844, 10.06, 16.57, 23.19, 21.32, 31.6, 11.83, 
18.68, 18.06, 15.4, 15.6, 19, 22.47, 17.36, 32.44, 7.074, 4.695, 8.94, 15.38, 
21.47, 18.29, 30.91, 18.62, 30.03, 37.07, 36.11, 33.96, 30.52, 30.18, 29.41, 
24.5, 37.03, 28.77, 20.77, 20.37, 18.03, 22.53, 13.21, 11.7, 15.23, 18.15, 
18.23, 80.89, 72.16, 69.33, 67.71, 72.56, 65.35, 62.66, 62.49, 67.48, 56.26, 
18.08, 6.25, 2.57, 4.85, 10.07, 16.58, 23.21, 21.34, 31.63, 9.03, 20.54, 27.04, 
26.76, 23.47, 24.48, 27.02, 29.21, 23.8, 38.76, 11.84, 7.08, 6.25, 4.7, 8.95, 
15.4, 21.49, 18.31, 30.95, 15.81, 27.25, 34.29, 33.35, 31.17, 27.96, 27.95, 
27.64, 22.55, 35.71, 39.54, 20.66, 27.95, 44.41, 36.14, 13.21, 22.53, 27.57, 
90.03, 32.43, 25.28, 93.22, 17.98, 65.35, 72.56, 79.16, 33.35, 38.03, 42.5, 
40.33, 50.4, 10.07, 4.85, 20.84, 10.65, 57.33, 65.38, 18.22, 73.56, 24.48, 
23.47, 10.82, 27.25, 41.4, 47.17, 34.39, 55.22, 8.95, 4.7, 15, 57.6, 67.19, 
7.74, 75.32, 27.96, 31.17, 12.8, 17.11, 41.23, 51.68, 19.79, 59.64, 17.68, 
27.32, 11.77, 19.45, 40.93, 49.07, 25.75, 57.26, 10.93, 14.8, 6.72, 18.94, 
38.67, 48.52, 22.98, 56.59, 2.96, 23.08, 10.13, 29.96, 30.72, 38.21, 35.54, 
46.4, 13.29, 14.93, 17.55, 60.03, 63.07, 27.32, 7.24, 21.1, 56.9, 61.45, 14.8, 
9.84, 8.51, 56.77, 60.25, 23.08, 3.61, 17.25, 45.99, 50.69, 14.93, 11.26, 
13.53, 57.44, 8.76, 9.02, 18.55, 12.94, 8.46, 53.95, 62.93, 5.64, 11.13, 10.68, 
5.25, 12.77, 59.17, 7.23, 55.63, 49.35, 70.49, 65.35, 47.34, 8.77, 17.84, 
51.18, 45.67, 65.47, 61.39, 44.94, 14.22, 58.09, 51.91, 73, 68.11, 50.22, 4.43, 
64.15, 59.19, 38.21, 49.19, 69.09, 61.83, 29.91, 55.27, 15.41, 29.83, 85.4, 
4.85, 17.38, 8.09, 75.63, 21.34, 8.39, 23.91, 86.58, 9.76, 70.92, 74.17, 57.97, 
87.6, 69.09, 17.92, 2.41, 56.55, 91.12, 60.99, 66.32, 51.15, 75.63, 61.83, 
6.05, 18.53, 50.78, 78.31, 22.57, 25.79, 37.71, 5.25, 29.91, 76.26, 88.96, 
40.47, 58.88, 60.51, 44.5, 76.47, 55.27, 23.23, 17.64, 42.5, 80.62, 65.58, 
68.11, 51.91, 82.76, 62.93, 19.2, 8.36, 50.22, 86.58, 9.36, 10.24, 6.29, 26.99, 
5.64, 52.41, 62.14, 8.72, 31.84, 15.15, 16.22, 33.01, 11.13, 47.01, 56.15, 3, 
37.71, 9.51, 5.61, 21.15, 15.4, 10.68, 66.57, 76.92, 23.22, 20.65, 8.72, 16.22, 
20.56, 5.25, 62.6, 72.32, 18.01, 25.79, 17.91, 18.01, 3, 35.69, 12.77, 46.37, 
54.8, 40.47, 61.52, 64.3, 48.09, 78.58, 59.17, 15.82, 9.29, 46.52, 82.35, 
13.29, 9.36, 17.19, 12.73, 55.02, 10.95, 9.34, 9.3, 16.93, 15.7, 8.72, 10.24, 
21.56, 15.03, 57.48, 14.48, 10.82, 5.25, 12.94, 28.31, 15.15, 6.29, 32.24, 
27.72, 41.28, 5.18, 23.91, 11.13, 9.02, 4.91, 17.88, 26.99, 4.09, 5.57, 72.3, 
28.74, 9.8, 33.07, 24.78, 19.87, 9.3, 5.64, 25.17, 19.21, 52.31, 10.3, 15, 8.3, 
69.83, 57.83, 52.41, 70.54, 69.46, 17.4, 48.17, 66.95, 54.69, 57.91, 81.64, 
68.91, 62.14, 83.15, 81.2, 16.14, 58.44, 78.24, 62.51, 67.28, 30.94, 17.91, 
8.72, 35.07, 30.33, 39.61, 8.18, 26.42, 8.46, 12.77, 10.1, 22.57, 31.84, 5.83, 
10.77, 76.22, 33.23, 15, 38.21, 29.91, 6.89, 7.84, 17.2, 9.39, 6.58, 61.65, 
18.3, 5.87, 24.47, 16.4, 12.26, 3.11, 10.02, 17.14, 11.63, 56.92, 12.72, 7.68, 
16.29, 8.1, 28.16, 16.93, 8.76, 33.42, 27.5, 46.95, 12.11, 23.28, 8.3, 37.33, 
24.12, 16.69, 40.19, 36.81, 30.97, 13.24, 33.4, 18.61, 21.92, 65.51, 53.29, 
47.61, 66.48, 65.12, 14.46, 43.44, 62.49, 53.07, 49.7, 68.12, 55.02, 47.47, 
70.4, 67.63, 44.2, 64.33, 46.95, 52.31, 19.98, 7.19, 2.24, 24.34, 19.38, 49.46, 
5.82, 15.48, 10.24, 4.91, 10.72, 9.51, 14.9, 16.98, 10.06, 62.37, 18.52, 6.14, 
18.55, 10.68, 61.87, 57.6, 48.6, 31.76, 10.8, 3.99, 50.34, 63.27, 41.19, 42.38, 
45.53, 30.22, 28.74, 40.33, 39.2, 48.78, 55.54, 63.27, 68.47, 39.3, 48.78, 
47.74, 8.72, 66.85, 5.61, 74.4, 15.7, 79.85, 47, 21.56, 8.72, 72.71, 47.6, 
81.83, 66.85, 10.29, 43.28, 76.52, 61.52, 12.03, 38.83, 26.54, 9.36, 62.48, 
50.8, 4.71, 250, 1, 250, 250, 250, 250, 250, 250, 250, 250, 250, 250, 250, 250, 
250, 250, 250, 250, 250, 250, 250, 250, 250, 250, 250, 250, 250, 250, 250, 1, 
250, 1, 2.27, 5.84, 0, 18.55, 5.75, 35.87, 26.88, 3.51, 8.46, 11.6, 18.95, 
50.07, 39.5, 16, 23.22, 24.48, 5.61, 13.69, 45.41, 35.09, 10.47, 18.01, 19.62, 
16, 4.14, 36.14, 26.5, 8.19, 10.71, 10.68, 39.5, 23.22, 5.25, 35.09, 18.01, 
5.36, 26.5, 8.19, 60.11, 30.21, 10.68, 33.16, 39.5, 47.34, 18.5, 23.22, 52.78, 
20.99, 18.52, 55.22, 26.5, 16, 14.06, 14.48, 30.2, 7.56, 10.47, 37.99, 250, 1, 
5.61, 10.47, 10.47, 250, 1, 0, 65.35, 71.71, 11.23, 11.27, 250, 250, 250, 250, 
250, 250, 250, 250, 250, 250, 250, 250, 250, 250, 250, 250, 250, 250, 250, 250, 
250, 250, 250, 250, 250, 250, 250, 250, 1, 250, 1, 250, 250, 250, 250, 1, 250, 
250, 250, 1, 3.49, 250, 1, 250, 250, 250, 1, 250, 1, 250, 250, 250, 250),
#startind variable#
startinds = c( 1, 2, 3, 4, 5, 6, 7, 8, 9, 10, 11, 12, 13, 14, 15, 16, 17, 
18, 19, 20, 21, 22, 23, 24, 25, 26, 27, 28, 29, 30, 31, 32, 33, 34, 35, 36, 37, 
38, 39, 40, 41, 42, 43, 44, 45, 46, 47, 48, 49, 50, 51, 52, 53, 54, 55, 56, 57, 
58, 59, 60, 61, 62, 63, 64, 65, 66, 67, 68, 69, 70, 71, 72, 73, 74, 75, 76, 77, 
78, 79, 80, 81, 82, 83, 84, 85, 86, 87, 88, 89, 90, 91, 92, 93, 94, 95, 98, 99, 
100, 101, 102, 103, 104, 105, 106, 107, 108, 109, 110, 111, 112, 113, 114, 115, 
116, 117, 118, 119, 120, 121, 122, 123, 124, 125, 126, 127, 128, 129, 130, 131, 
132, 133, 134, 135, 136, 137, 138, 139, 140, 141, 142, 144, 145, 152, 197, 255, 
302, 334, 354, 408, 503, 591, 669, 685, 691, 697, 705, 713, 715, 716, 717, 718, 
719, 720, 721, 722, 723, 724, 725, 726, 727, 728, 729, 730, 731, 732, 733, 734, 
735, 736, 737, 738, 739, 740, 741, 742, 743, 744, 745, 746, 747, 749, 750, 757, 
776, 785, 793, 799, 805, 806, 807, 809, 810, 811, 812, 813, 814, 815, 817, 818, 
819, 820, 821, 822, 823, 824, 825, 826, 827, 828, 829, 830, 831, 832, 833, 834, 
835, 836, 837, 838, 839, 840, 841, 842, 843, 844, 845, 846, 847, 848, 849, 850, 
851, 852, 853, 854, 855, 856, 857, 858, 859, 860, 861, 862, 863, 864, 865, 866, 
867, 868, 869),
#endind variable#
endinds = c( 1, 2, 3, 4, 5, 6, 7, 8, 9, 10, 11, 12, 13, 14, 15, 
16, 17, 18, 19, 20, 21, 22, 23, 24, 25, 26, 27, 28, 29, 30, 31, 32, 33, 34, 35, 
36, 37, 38, 39, 40, 41, 42, 43, 44, 45, 46, 47, 48, 49, 50, 51, 52, 53, 54, 55, 
56, 57, 58, 59, 60, 61, 62, 63, 64, 65, 66, 67, 68, 69, 70, 71, 72, 73, 74, 75, 
76, 77, 78, 79, 80, 81, 82, 83, 84, 85, 86, 87, 88, 89, 90, 91, 92, 93, 94, 97, 
98, 99, 100, 101, 102, 103, 104, 105, 106, 107, 108, 109, 110, 111, 112, 113, 
114, 115, 116, 117, 118, 119, 120, 121, 122, 123, 124, 125, 126, 127, 128, 129, 
130, 131, 132, 133, 134, 135, 136, 137, 138, 139, 140, 141, 143, 144, 151, 196, 
254, 301, 333, 353, 407, 502, 590, 668, 684, 690, 696, 704, 712, 714, 715, 716, 
717, 718, 719, 720, 721, 722, 723, 724, 725, 726, 727, 728, 729, 730, 731, 732, 
733, 734, 735, 736, 737, 738, 739, 740, 741, 742, 743, 744, 745, 746, 748, 749, 
756, 775, 784, 792, 798, 804, 805, 806, 808, 809, 810, 811, 812, 813, 814, 816, 
817, 818, 819, 820, 821, 822, 823, 824, 825, 826, 827, 828, 829, 830, 831, 832, 
833, 834, 835, 836, 837, 838, 839, 840, 841, 842, 843, 844, 845, 846, 847, 848, 
849, 850, 851, 852, 853, 854, 855, 856, 857, 858, 859, 860, 861, 862, 863, 864, 
865, 866, 867, 868, 869))

    \end{verbatim}  
\endgroup

\end{document}